\documentclass[prd,nofootinbib,superscriptaddress,showpacs,twocolumn,reprint]{revtex4}
\usepackage[a4paper, total={6.5in, 8.5in}]{geometry}
\usepackage{graphicx}
\usepackage{bm}
\usepackage[usenames,dvipsnames]{color}
\usepackage{amsmath,amssymb}
\usepackage{slashed}
\usepackage{float}
\usepackage{epsfig}
\usepackage{latexsym, amssymb} 
\usepackage[colorlinks=true,linkcolor=blue,citecolor=blue]{hyperref}
\usepackage{IEEEtrantools}
\begin{document}
	
\title{Scalar Induced Gravitational Waves from Warm Inflation} 	
\author{Richa Arya}
\email{richaarya@iisc.ac.in} 
\affiliation{
Indian Institute of Science, Bangalore 560002, India}
\author{Arvind Kumar Mishra} 
\email{arvind.mishra@acads.iiserpune.ac.in}
\affiliation{Indian Institute of Science Education and Research, Pune 411008, India}

\date{\today}

\begin{abstract}
Stochastic gravitational waves can be induced from the primordial curvature perturbations generated  during inflation, through scalar-tensor mode coupling at the second order of cosmological  perturbation theory. Here we discuss a model of warm inflation in which large curvature perturbations are generated at the small scales because of inflaton dissipation. 
These overdense perturbations then collapse at later epoch to form primordial black holes, as was studied in our earlier work (Ref. \cite{Arya:2019wck}), and therefore may also act as a source to the second-order tensor perturbations. In this study, we calculate the spectrum of these secondary gravitational waves from our warm inflationary model. We find that our model leads to a generation of scalar induced gravitational waves (SIGW) over a frequency range ($1-10^6$) Hz. Further, we discuss the detection possibilities of these SIGWs, taking in account the sensitivity of different ongoing and future gravitational wave experiments.  
\end{abstract}

\maketitle
\newpage

\section{Introduction}
The inflationary paradigm of the early Universe is a widely well-known phase of a rapid accelerated expansion, which resolves many issues of the Standard Model of cosmology \cite{Kazanas:1980tx,Sato:1980yn,Guth:1980zm,Starobinsky:1980te,Linde:1981mu}. Additionally, it also generates  density perturbations for all the structures  we see today. During inflation, the primordial (scalar, vector, tensor) perturbations  of a vast range of wavelengths are generated. According to the decomposition theorem, at the linear order of cosmological perturbation theory, these perturbation modes
evolve independently. The scalar fluctuations are imprinted as the temperature anisotropies in the Cosmic Microwave Background (CMB) radiation and further become the seed density perturbations that grow into large-scale structures (LSS) at the late time. 
On the large scales $10^{-4}$ Mpc$^{-1}$ $<k<1$ Mpc$^{-1}$, the amplitude of the primordial scalar curvature perturbations has been precisely measured 
to a value $\Delta^2_\mathcal{R} (k_P)=2.1\times 10^{-9}$ \cite{Akrami:2018odb}. The tensor perturbations signature on the CMB as B-mode polarization have not been detected till date. However 
an upper limit has been imposed on the amplitude of the primordial tensor fluctuations, through the observable tensor-to-scalar ratio $r$. The latest observations of BICEP/Keck have put a constraint on $r<0.036$ at 95$ \%$ confidence level \cite{BICEP:2021xfz}. 
A detection of primordial gravitational waves would provide information about the energy scale of inflation and thus is important to understand 
the physics of early Universe.
For a review on inflationary cosmology, see Refs. \cite{Baumann:2009ds,Linde:2007fr,Tsujikawa:2003jp,Olive:1989nu,Riotto:2002yw}.

Besides the linear theory, the scalar perturbations can also couple to the tensor perturbations and generate secondary gravitational waves at the second order of cosmological perturbation theory \cite{Mollerach:2003nq,Ananda:2006af,Baumann:2007zm}. 
These can be inevitably seen in the scenario of primordial black hole (PBH) production from inflationary models \cite{Saito:2008jc,2010PThPh.123..867S}. 
Primordial Black Holes  \cite{zeldo:1966,Hawking:1971ei,Carr:1974nx,Carr:1975qj}, i.e. black holes with a primordial origin, serve as a probe to a huge range of small scales,  
unexplored by the CMB and LSS observations. 
They can form in a number of ways - by the collapse of overdense fluctuations \cite{Hawking:1971ei,Carr:1974nx}, from the collision of bubbles \cite{Hawking:1982ga}, by the collapse of strings \cite{Hogan:1984zb,Hawking:1987bn} or domain walls \cite{Caldwell:1996pt}, etc. As for PBH generation, 
the amplitude of the primordial curvature power spectrum at small scales is hugely amplified $\mathcal{O}(10^{-2})$, it is quite evident that these large scalar perturbations then 
act as a source term to the second-order tensor fluctuations.
In literature, 
the generation of primordial black holes and the associated secondary gravitational waves has been
investigated for many inflationary models, for
example, runnning mass inflation  \cite{2010PhRvD..81b3517B,Alabidi:2012ex}, double inflation \cite{Inomata:2018cht}, axion curvaton model \cite{Orlofsky:2016vbd}, hybrid inflation with chaotic potentials \cite{Ahmed:2021ucx}, 
inflation with gravitationally enhanced friction \cite{Fu:2019vqc}, chaotic inflation \cite{Gao:2021vxb}, models with noncanonical kinetic term \cite{Yi:2020cut,Lin:2021vwc}, inflection point models \cite{Bhaumik:2020dor}, 
 ultraslow roll and punctuated inflation  \cite{Ragavendra:2020sop}, $k$ and $G$ inflation\cite{Lin:2020goi}, etc.
The spectrum of these scalar induced gravitational waves (SIGW) can be constrained through various ongoing and future gravitational wave experiments  \cite{2009Natur.460..990A,Assadullahi:2009jc,Inomata:2018epa}, such as Pulsar Timing Array (PTA \cite{Lentati:2015qwp}, NANOGrav \cite{NANOGrav:2015aud}, and SKA \cite{Janssen:2014dka}), second-generation GW interferometers (advanced LIGO, VIRGO, KAGRA \cite{KAGRA:2013rdx}), space based GW interferometers (LISA \cite{2017arXiv170200786A}, DECIGO, BBO \cite{Yagi:2011wg}), third-generation GW interferometers (Einstein Telescope (ET) \cite{ET}, Cosmic Explorer (CE) \cite{2015PhRvD..91h2001D}). For a review, see Refs. \cite{Domenech:2021ztg,Bian:2021ini}.

In this study, we consider a scenario, known as  \textit{warm inflation}\cite{Berera:1995wh,Berera:1995ie,Berera:1998px},
 in which the dynamics of the Universe is governed by the dissipative and non-equilibrium effects in a coupled inflaton-radiation system. In this description, the inflaton interactions to the other fields are not neglected during inflation, unlike in the standard cold inflation, and there is a non-zero temperature in the Universe throughout the inflationary phase. The background evolution as well as the perturbations  of the inflaton field are modified due to its coupling with other fields. As a result, the primordial power spectrum and thus the predicted observables in warm inflation differ from those in cold inflation. (For a review, see Refs. \cite{Berera:2006xq,Berera:2008ar}.) Warm inflation is well-motivated for multiple reasons, as will be discussed in Section \ref{Sec2}.  In the literature, many warm inflationary models are constrained in the context of CMB observations \cite{Bastero-Gil:2015nja,Visinelli:2016rhn,Benetti:2016jhf,Arya:2017zlb,Bastero-Gil:2017wwl,Arya:2018sgw}, however the small scale features of warm inflation are not well explored. Recently,  we had studied the formation of primordial black holes during radiation dominated era, in  a model of warm inflation \cite{Arya:2019wck}.  
 We found that for some parameter space, we can successfully explain the CMB on large scales as well as generate light mass PBHs (nearly $10^3$ g) of a significant abundance.  
 More recently, the authors of Ref. \cite{Bastero-Gil:2021fac} carried out a study on the secondary gravity waves and PBH production in a scalar warm little inflaton model. 
 They found that PBHs with mass lighter than $10^6$ g and a spectrum of GW peaked at $10^5-10^6$ Hz could be generated in their model. 
 
 In this work, we aim to discuss the secondary gravitational waves induced by the large scalar curvature perturbations in warm inflation. To this extent, we consider a quartic potential ($V(\phi)=\lambda \phi^4$) and a dissipation coefficient cubicly dependent on temperature ($\Upsilon\propto T^3)$.  The motivation of considering this potential is that it is the simplest single-field warm inflation model which, for some parameter space, is consistent with the CMB observations and also predicts a large amplitude of scalar power spectrum on the very small scales, leading to PBH formation.  Further, as the second order tensor modes are coupled with the first order scalar modes, we expect a spectrum of secondary gravitational  wave induced by the large scalar fluctuations. Indeed, we find that our warm inflationay model leads to a GW spectrum with a significant amplitude over the frequency range $1-10^6$ Hz, which may be explored in the future high frequency detectors, such as the
 levitated-sensor detector \cite{Aggarwal:2020umq}, microwave cavities \cite{Bernard:2002ci}, decameter Michelson interferometers \cite{Holometer:2016qoh}, resonant mass detectors {\cite{Aguiar:2010kn}}. We will discuss it in detail in Section \ref{Sec4}.

This paper is organised as follows: In Section \ref{Sec2}, we discuss the basics of warm inflation theory, the background evolution of inflaton, dissipation coefficient and the primordial curvature power spectrum of warm inflation. 
 Then in Section \ref{Sec3}, we discuss the spectrum of induced gravitational waves sourced by the scalar curvature perturbations. After this, we discuss our model of warm inflation and the results obtained in Section \ref{Sec4} and then finally summarise the paper in Section \ref{Sec5}. We also present an Appendix \ref{A} for the calculation of dissipation coefficient in warm inflation.

\section{Warm Inflation}
\label{Sec2}
Warm Inflation 
 is a description of inflation in which one considers dissipative processes during the inflationary phase. In contrast to the standard cold inflation, where one neglects the inflaton couplings to other fields during the slow-roll inflationary phase, warm inflation is a more general and natural description, where the inflaton interactions to other fields are considered. In this picture, the inflaton dissipates its energy into radiation fields, and as a result of which there is a non-zero temperature in the Universe during the inflationary phase. Depending on the strength of inflaton dissipation, warm inflation is classified into two dissipative regimes - weak and strong, characterized by a dissipation parameter $Q$ defined as the ratio of the inflaton dissipation rate to the Hubble expansion rate.

Warm inflation is motivated for multiple reasons: At first, it is a more complete picture of inflation and has cold inflation as its limiting case. As particle production takes place simultaneously with the expansion during warm inflation,  a separate reheating phase may not be required in some models \cite{Berera:1996fm}. Warm inflation predicts unique and distinct characteristics from cold inflation on the CMB.  Certain models of cold inflation which are ruled out from observational constraints, are viable models of warm inflation for some range of dissipation \cite{Visinelli:2016rhn,Benetti:2016jhf,Arya:2017zlb,Bastero-Gil:2017wwl,Arya:2018sgw}. 
Besides the Gaussian two-point correlations, warm inflation can also lead to non-Gaussianities because of the inflaton interactions
and hence can be tested in
the future CMB experiments \cite{Gupta:2002kn,Moss:2011qc}. 
In warm inflation, the conditions for slow roll are modified due to an extra friction term in the
inflaton equation of motion. Thus, 
warm inflation is less restrictive in the flatness of the potential, and thus may relax the $\eta$ problem \cite{Berera:2004vm}. 

Further, warm inflation also predicts interesting features at the small scales. In Ref. \cite{Arya:2019wck}, we found that on the small scales, the primordial power spectrum is blue-tilted ($n_s>1$) and has a large amplitude, which leads to the formation of primordial black holes.
Furthermore, as inflation is a low energy effective
field theory, it has to obey some criteria, such as the swampland distance and de-Sitter
conjectures, in order to embed it in a UV complete theory. It has been
found that single-field slow roll cold inflation, with a canonical kinetic term
and a Bunch Davies vacuum, is not in accordance with the swampland conjectures.
However, recent studies show that warm inflation  models 
with a large value of the dissipation parameter  \cite{Das:2018rpg,Motaharfar:2018zyb}
can satisfy the swampland conjectures, thus making it in agreement with a high energy theory.
All these above features arise from the fundamental feature of treating the dynamics
of inflaton as that of a dissipative system, and hence makes warm inflation interesting.
Thus, a comprehensive study of the warm inflation scenario is  necessary
to understand the physics of the early Universe.

\subsection{Background evolution equations}
In the warm inflation description, one
considers the dissipative processes during inflation based on the principles
of non-equilibrium field theory for interacting quantum fields \cite{PhysRevD.37.2878,Gleiser:1993ea}. 
The inflaton is assumed to be near-equilibrium and evolving slowly as compared to the microphysics timescales in the adiabatic approximation.
The effective equation of motion of inflaton field 
is obtained using the Schwinger-Keldysh close time path formalism of thermal field theory. (For a review, see Refs. \cite{Das:1997gg,lebellac:1996}).
Due to its couplings, the equation of motion 
is modified by an additional friction term and is given by \cite{Gleiser:1993ea,Berera:1998gx}
\begin{equation}
\ddot \phi(t) + 3 H \dot\phi(t) +  \Upsilon \dot\phi(t) + V'(\phi)= 0.
\label{eominf}
\end{equation}
Here $\Upsilon(\phi, T)$ is the dissipation coefficient, which depends on the mechanism of inflaton dissipation, such as the channel of decay, the coupling strengths, and the multiplicities of the fields involved. We define a dissipation parameter $Q=\Upsilon/3H$ and rewrite Eq. (\ref{eominf}) as
\begin{equation}
\ddot \phi(t) + 3 H (1+Q) \dot\phi(t) + V'(\phi)= 0.
\label{eominfQ}
\end{equation}
When the dissipation parameter is smaller than the Hubble
expansion rate ($Q < 1$), it is the weak dissipative regime, and when it is larger ($Q > 1$),
then it is the strong dissipative regime of warm inflation.

From the continuity equation, we can also write the energy
density of the inflaton $\rho_\phi$ and radiation $\rho_r$ as,
 \begin{equation}
 \dot{\rho_\phi}+3H ({p_\phi+\rho_\phi})=-\Upsilon \dot{\phi}^2.
 \label{rhophi}
 \end{equation}
 \begin{equation}
  \dot\rho_r+4H\rho_r=\Upsilon{\dot\phi}^2~.
  \label{rad}
  \end{equation}
 The negative sign on the right-hand side of Eq. (\ref{rhophi}) shows that the inflaton dissipates its energy with time.
 As a result of the dissipation, radiation is 
 produced along with the expansion during warm inflation, as can be seen on Eq. (\ref{rad}).
 Assuming that the radiation thermalizes  quickly after being produced, we have $\rho_r=\frac{\pi^2}{30} g_* T^4$ where $T$ is the temperature of the thermal bath, $g_*$ is the number of relativistic degrees of freedom present during warm inflation.
 
 \subsubsection{Slow roll parameters and conditions}
 The flatness of the potential $V(\phi)$ in inflation is measured in terms of the potential slow roll parameters, similar to the ones defined for cold inflation 
 \begin{equation}
 \epsilon_\phi = 
 \frac{M_{Pl}^2}{16\pi}\,\left(\frac{V'}{V}\right)^2, \hspace{1cm}
 \eta_\phi = 
 \frac{M_{Pl}^2}{8\pi}
 \,\left(\frac{V''}{V}\right).
 \end{equation}
 In addition to these, in warm inflation there are other slow roll parameters defined as \cite{Hall:2003zp, Moss:2008yb}
 
 \begin{equation}
 \beta_\Upsilon = 
 \frac{M_{Pl}^2}{8\pi}
 \,\left(\frac{\Upsilon'\,V'}{\Upsilon\,V}\right),  \hspace{0.5cm}
 b=\frac{T V'_{,T}}{V'}~,\hspace{0.5cm} c=\frac{T\Upsilon_{,T}}{\Upsilon}.
 \label{slowroll}
 \end{equation}
 Here the subscript $_{,T}$ represents derivative of the quantity w.r.t $T$. These additional slow roll parameters are a measure of the field and temperature dependence in the inflaton potential and the dissipation coefficient. The stability analysis of warm inflationary solution shows that the following conditions should be satisfied during the slow roll \cite{Moss:2008yb} 
 \begin{align} 
 &\epsilon_\phi \ll 1+Q,\hspace{1cm} |\eta_\phi| \ll 1+Q,\hspace{0.5cm} |\beta_\Upsilon| \ll 1+Q, 
 \nonumber\\
 &0<b\ll\frac{Q}{1+Q}, \hspace{1cm} \hspace{0.5cm} |c|<4.
 \label{slow_roll}
 \end{align}
 As can be clearly seen, for large $Q$, these  conditions relax the requirement for the potential to be extremely flat, as the upper limit on the slow roll parameters $\epsilon_\phi, \eta_\phi$ is increased.
 Therefore, the $\eta$ problem is not as severe in warm inflation.
 
 \subsubsection{Evolution equations in the slow roll approximation}
 In the slow roll approximation, we can neglect $\ddot \phi$ in Eq. (\ref{eominfQ}) which gives
 \begin{equation}
 \dot\phi\approx \frac{-V'(\phi)}{3H(1+Q)},
 \label{phido}
 \end{equation}
and since $\dot\rho_r$ is smaller than the other terms in Eq. (\ref{rad}) throughout inflation, we can approximate $\dot\rho_r\approx 0$ and obtain 
\begin{equation}
\rho_r 
\approx \frac{\Upsilon}{4H} {\dot\phi}^2 =\frac{3}{4} Q {\dot\phi}^2.
\label{rhodot}
\end{equation}

\subsection{Dissipation Coefficient}
\label{dc}
The microphysics of the coupled inflaton-radiation system 
 results into a dissipation coefficient in the inflaton equation of motion. 
Depending on the interaction Lagrangian, the channel of inflaton decay, the coupling strengths, and the multiplicities of the fields involved, there are different model constructions of warm inflation.
 In the earlier ones, it was realized that it is difficult to obtain a successful strong dissipative regime, as the thermal corrections to the effective potential
 are large
 \cite{Berera:1998gx,Yokoyama:1998ju}. Therefore, subsequent studies considered models, such as the supersymmetric distributed mass model in the context of string theory \cite{Berera:1998px,Bastero-Gil:2018yen}, or a two-stage
 decay mechanism of inflaton, where the inflaton couples to a heavy intermediate catalyst field which then further couple to the light radiation fields \cite{Berera:2001gs,Berera:2004kc}
 or recently discrete interchange symmetry in the warm little inflaton model \cite{Bastero-Gil:2016qru,BASTEROGIL2021136055}
 to control these corrections, and attain a strong dissipation regime of warm inflation. Here we will
consider a two-stage decay of the inflaton in a supersymmetric inflation model \cite{Moss:2006gt,BasteroGil:2010pb}. 
In this, we have three superfields $\Phi$, $ X$, and $Y$, whose scalar and fermion components are ($\phi$, $\psi_\phi$), ($\chi$,$\psi_\chi$) and ($\sigma$,$\psi_\sigma$), respectively.
The interacting superpotential is given as
 \begin{equation}
 W=g\Phi X^2 +h X Y^2,
 \label{super}
 \end{equation}
 where $g$ and $h$ are the coupling strengths between $\Phi-X$, and $X-Y$, respectively. 
 The scalar inflaton is coupled with the intermediate bosonic and fermionic components of the $X$ superfield (also called catalyst fields), which subsequently decay into the scalar and fermionic components of the $Y$ superfield (called radiation fields).  The radiation fields are considered to be lighter than the catalyst fields. The scatterings of decay products $\sigma,\psi_\sigma$  with masses $m_\sigma,m_{\psi_\sigma}\ll T$ is sufficient to keep them thermalized  and constitute the thermal bath, as shown in the Appendix C of Ref.  \cite{BasteroGil:2012cm}.  The inflaton particle states are also assumed to thermalize with a same temperature, for some range of effective couplings \cite{BasteroGil:2012cm}.
  The scalar part of the Lagrangian is given as
  	\begin{align}
  	-\mathcal{L}_s & = |\partial_\Phi W|^2+|\partial_{{X}} W|^2+|\partial_{{Y}} W|^2 \nonumber \\
  	& = g^2|\chi|^4+h^2|\sigma|^4+4g^2|\chi|^2|\phi|^2\nonumber \\
  	 & +4gh Re[\phi^\dagger\chi^\dagger\sigma^2]+4 h^2 |\chi|^2|\sigma|^2.
  	\label{scalar}
  	\end{align}
  	The scalar fields $\phi,\chi,\sigma$ are chosen to be complex,  
  	$\chi=(\chi_1+i\chi_2)/\sqrt{2}$, and similarly for others.
  When the background scalar field $\phi$ takes an expectation value $\mathcal{\varphi}/\sqrt{2}$, the mass of the $\chi$ field is given as:
  $m_{\chi_1}=\sqrt{2}g\varphi, m_{\chi_2}=\sqrt{2}g\varphi.$	
 The Yukawa interactions are obtained as
  	\begin{equation}
  	-\mathcal{L}_Y=\frac{1}{2}\sum_{n,m}^{}\frac{\partial^2 W}{\partial \zeta_n~\partial \zeta_m}\bar\psi_n P_L \psi_m + \frac{1}{2}\sum_{n,m}^{}\frac{\partial^2 W^\dagger}{\partial \zeta_n^\dagger~\partial \zeta_m^\dagger}\bar\psi_n P_R \psi_m
  	\end{equation}
  	where $\zeta$ refers to the superfields $\Phi,X,Y$ and $P_L=1-P_R=(1+\gamma_5)/2$. For the superpotential given in Eq. (\ref{super}), the Yukawa interactions are given as
  		\begin{align}
  		-\mathcal{L}_Y=&g\phi \bar{\psi}_\chi P_L\psi_\chi+2g\chi \bar{\psi}_\phi P_L\psi_\chi+h\chi \bar{\psi}_\sigma P_L\psi_\sigma \nonumber \\ &+2h\sigma \bar{\psi}_\chi P_L\psi_\sigma+ h.c.
  		\label{yukawa}
  		\end{align}	
 For these interaction terms, the dissipation coefficient is calculated, as shown in Appendix \ref{A}. In this study, we choose the special case, when the intermediate catalyst fields are heavy, which gives $\Upsilon=C_\phi T^3/\phi^2$ in the low temperature limit. 
 \subsection{Primordial curvature power spectrum}
 \label{primpower}
 In warm inflation description, as there is a temperature in the Universe throughout the inflationary phase,  therefore the fluctuations in the inflaton field are also sourced by  the thermal noise, unlike in the cold inflation where the inflaton has only quantum fluctuations. 
 The total primordial curvature power spectrum for warm inflation by including both quantum and thermal contributions to the inflaton power spectrum
 is given as  \cite{Hall:2003zp,Graham:2009bf,BasteroGil:2011xd,Ramos:2013nsa,Bartrum:2013fia,Bastero-Gil:2016qru,Benetti:2016jhf} 
 \begin{align}
 \Delta^2_\mathcal{R}(k)=&\left(\frac{H_k^2}{2\pi\dot\phi_k}\right)^2\left[~1+2n_k+\left(\frac{T_k}{H_k}\right)
 \frac{2\sqrt{3}\pi Q_k}{\sqrt{3+4\pi Q_k}}~\right]\nonumber \\
 &\times G(Q_k).
 \label{power}
 \end{align}
Here is the description of each term present in this equation:
 \begin{itemize}
 	\item The prefactor $\left(\frac{H_k^2}{2\pi\dot\phi_k}\right)^2$
 	is the primordial curvature power spectrum in the cold inflation.
 	It shows that in the limit  $Q\rightarrow 0$ and $ T\rightarrow 0,$ we recover the standard cold inflation from warm inflation.
 	\item 
 Due to the presence of the radiation bath in warm inflation, the inflaton can also be excited
 from its vacuum state to  
 Bose-Einstein distribution, given as
$
 n_k=\frac{1}{\exp(\frac{k/a_k}{T_k}) -1}.
$
The system of inflaton particles and radiation fields is assumed to thermalize with a same temperature, and the scattering rates are shown in Ref. \cite{BasteroGil:2012cm}.
 
  \item Due to the thermal noise contributions to the inflaton fluctuations, 
 the primordial power spectrum has terms dependent on the dissipation coefficient and the temperature of the thermal bath, as given by the third term in the square bracket.
 \item The perturbations in the radiation 
 can also couple to the  inflaton perturbations and lead to a growth in the primordial power spectrum \cite{Graham:2009bf}. This growth factor $G(Q_k)$ depends on the form of dissipation coefficient and is 
 obtained numerically \cite{Bastero-Gil:2016qru,Benetti:2016jhf}. For the form of dissipation coefficient under consideration
  $\Upsilon \propto T^3,$ 
  \begin{equation}
  G(Q_k)= 1+4.981 \,Q_k^{1.946} +0.127 \,Q_k^{4.330}.
  \label{GQ}
  \end{equation}
  In the weak dissipation regime, the growth factor does not enhance the power spectrum significantly. But in the strong dissipation regime, the power spectrum is considerably enhanced due to the growth factor. 
  \end{itemize}
  Further, it is interesting to note that in the strong dissipation regime, the shear effects in radiation also become important which cause damping of the power spectrum \cite{BasteroGil:2011xd}, and therefore the overall growth in the power spectrum is reduced. In the expression for the primordial power spectrum given above, we do not account for any shear effects.

 \section{Scalar Induced Gravitational Waves spectrum}
 \label{Sec3}
 In this Section, we briefly review the gravitational waves spectrum induced from the scalar perturbations at second order of cosmological perturbation theory. For a more detailed derivation, we suggest Refs. \cite{Ananda:2006af,Baumann:2007zm,Kohri:2018awv}. To start with, we consider a perturbed metric in the longitudinal gauge with vanishing vector perturbations as
 \begin{equation}
 ds^{2}=-a^{2}\left(1+2\Phi \right) d\eta^{2}+a^{2}\left[\left(1-2\Psi \right)\delta_{ij} +\frac{h_{ij}}{2}  \right]dx^{i}dx^{j}
 \end{equation}
 where $a$ is the scale factor and $\eta$ is the conformal time. Here $\Phi$, $\Psi$ are the scalar and $h_{ij}$ correspond to the tensor metric perturbations, respectively. In our notation, the spatial coordinates $i,j,k,$ etc. can take values $1,2,3$. In this work, we focus our analysis for vanishing anisotropic stress for which $\Phi=\Psi$. However, in Ref. \cite{Baumann:2007zm}, it has been found that effect of anisotropic stress, i.e., $\Phi\neq\Psi$, is very small.

 The second order action for graviton can be given as 
 \begin{equation}
 S=\frac{\mathrm{M^{2}_{Pl}}}{32}\int d^{3}x ~ d\eta~ a^{2} \left( h'_{ij} h'_{ij}- h_{ij,k} h_{ij,k}\right)
 \end{equation} 
 where $\mathrm{M_{Pl}}=1/8\pi G$ is the reduced Planck mass, $h_{ij,k}$ represents derivative of $h_{ij}$ w.r.t. spatial coordinate and $h_{ij}'$ is the differentiation w.r.t. conformal time. We do a Fourier decomposition of 
 tensor $h_{ij}(\eta,\textbf{x})$ as
 \begin{equation}
 h_{ij}(\eta,\textbf{x})=\int \frac{d^{3}\mathrm{k}}{(2\pi)^{\frac{3}{2}}}\left[ e^{+}_{ij}(\textbf{k})h_{\textbf{k}}^{+}(\eta)+e^{\times}_{ij}(\textbf{k})h_{\textbf{k}}^{\times}(\eta)\right]e^{i\textbf{k}\cdot\textbf{x}}
 \end{equation}
 where $e^{+}_{ij}(\textbf{k})$ and $e^{\times}_{ij}(\textbf{k})$ are time independent, traceless, transverse vectors. These quantities are defined in terms of orthonormal basis, $e_{i}(\textbf{k})$, $\bar{e}_{i}(\textbf{k})$ as
 \begin{equation}
 e^{+}_{ij}(\textbf{k})= \frac{1}{\sqrt{2}} \left[e_{i}(\textbf{k}) e_{j}(\textbf{k})-\bar{e}_{i}(\textbf{k}) \bar{e}_{j}(\textbf{k})\right] 
 \end{equation}
 \begin{equation}
 \mathrm{and} \quad e^{\times}_{ij}(\textbf{k})= \frac{1}{\sqrt{2}} \left[e_{i}(\textbf{k}) \bar{e}_{j}(\textbf{k})+\bar{e}_{i}(\textbf{k}) e_{j}(\textbf{k})\right].
 \end{equation}
 Further, the power spectrum of tensor perturbation is defined as
 \begin{equation}
 \langle h_{\textbf{k}}^{\lambda}(\eta)h_{\textbf{k}^{\prime}}^{\lambda'}(\eta) \rangle = \frac{2\pi^{2}}{k^{3}} \delta_{\lambda\lambda'}\delta^{3}(\textbf{k}+\textbf{k}^{\prime})  \mathcal{P}_{h}(\eta,k)
 \label{eq:hpowdef}
 \end{equation}
 where, $\lambda,\lambda'= +, \times$ corresponds to the polarization index and $\mathcal{P}_{h}(\eta,k)$ is the 
 dimensionless tensor power spectrum. In our caculation, we assume parity invariance, which provides the same result for both the polarizations.
 
 To obtain $P_h(\eta, k)$, we first need to explore the dynamics of tensor mode, which can be obtained by using the Einstein's equation. The evolution of tensor mode sourced by the scalar perturbation is given as \cite{Kohri:2018awv}
 \begin{equation}
 h''_{\textbf{k}}(\eta) + 2\mathcal{H}h'_{\textbf{k}}(\eta) + \mathrm{k}^{2}h_{\textbf{k}}(\eta) = 4S_{\textbf{k}}(\eta)
 \label{eq:tensorevol}
 \end{equation}
 where $\mathcal{H}$ is the conformal Hubble parameter, and $S_{\textbf{k}}(\eta)$ is the source term  
 with quadratic contributions from scalar perturbations given by 
 \begin{align}
 S_{\textbf{k}}(\eta)& = \int \frac{d^{3}q}{(2\pi)^{\frac{3}{2}}} e_{ij}(\textbf{k}) q_{i}q_{j}\left[2\Phi_{\textbf{q}}\Phi_{\textbf{k-q}}\right.
 \nonumber  \\  & \left. 
 +\frac{4}{3(1+w)}\left(\mathcal{H}^{-1} \Phi'_{\textbf{q}}+\Phi_{\textbf{q}}\right) \left( \mathcal{H}^{-1} \Phi'_{\textbf{k-q}}+\Phi_{\textbf{k-q}} \right)  \right].
 \label{eq:ssol}
 \end{align}
 Further, to estimate $\langle h_{\textbf{k}}^{\lambda}(\eta)h_{\textbf{k}^{\prime}}^{\lambda'}(\eta) \rangle$, one needs to evaluate $\langle S_{\textbf{k}}(\eta)S_{\textbf{k}^{\prime}}(\eta') \rangle$, which in turn requires  the dynamics of potential field $\Phi$.  The evolution of $\Phi$ is calulated using the Einstein's equation and is given as \cite{Kohri:2018awv}
 \begin{align}
 \Phi''_{\textbf{k}} + 3\mathcal{H}(1+c^{2}_{\mathrm{s}})\Phi'_{\textbf{k}} &+ \left[ 2\mathcal{H}' +  (1+ 3c^{2}_{\mathrm{s}})\mathcal{H}^{2}+ c^{2}_{\mathrm{s}}\mathrm{k}^{2} \right]\Phi_{\textbf{k}} \nonumber \\ & = \frac{a^{2}}{2}\tau \delta S
 \end{align}
 where the $c^{2}_{\mathrm{s}}=\left(\frac{\delta p}{\delta \rho}\right)_S$ is the square of the speed of sound and $\delta S$ is the entropy perturbation.  The pressure perturbation can be expressed as $\delta P = c^{2}_{\mathrm{s}} \delta \rho + \tau \delta S$. To simplify the analysis,  we  assume that $ \delta S=0$ and sound speed is constant, $ c^{2}_{\mathrm{s}}=w$. Further, we parameterise the scalar field as  $\Phi_{\textbf{k}}=\Phi(\mathrm{k}\eta) \phi_{\textbf{k}}$, where $\phi_{\textbf{k}}$ is its primordial fluctuation and $\Phi(\mathrm{k}\eta)$ correpond to the transfer function. The two point correlation function for primordial fluctuation is given as   
 \begin{equation}
 \langle \phi_{\textbf{k}}\phi_{\textbf{k}'} \rangle = \frac{2\pi^{2}}{k^{3}} \delta^{3}(\textbf{k}+\textbf{k}^{\prime}) \left(\frac{3+3w}{5+3w} \right)^{2}   \mathcal{P}_{\zeta}(\mathrm{k})
 \end{equation}
 where $\mathcal{P}_{\zeta}(\mathrm{k})$ is the primordial curvature perturbations. 
Equipped with this expression, we will next obtain $\mathcal{P}_{h}(\eta,k).$

 The solution for $h_{\textbf{k}}(\eta)$ can be obtained by applying the Green's function method in Eq. (\ref{eq:tensorevol}) as
 \begin{equation}
 h_{\textbf{k}}(\eta)=\frac{4}{a(\eta)}\int d\tilde{\eta} ~G_{\textbf{k}}(\eta,\tilde{\eta})~ a(\tilde{\eta})~S_{\textbf{k}}(\tilde{\eta})
 \label{eq:hsol}
 \end{equation}
 where $G_{\textbf{k}}(\eta,\tilde{\eta})$ is the solution of differential equation
 \begin{equation}
 G''_{\textbf{k}}(\eta,\tilde{\eta}) + \left[\mathrm{k}^{2}-\frac{a''(\eta)}{a(\eta)} \right]G_{\textbf{k}}(\eta,\tilde{\eta})=\delta(\eta-\tilde{\eta}).
 \end{equation}
 Then, using Eq. (\ref{eq:hpowdef}) and Eq. (\ref{eq:hsol}), we get \footnote{In the r.h.s. of this  obtained equation, there is a discrepancy of a factor of $16$ in Ref. \cite{Baumann:2007zm} and  Ref. \cite{Kohri:2018awv}. This is because the source term in the evolution equation of $h_{\textbf{k}}$ in Eq. (\ref{eq:tensorevol}) are differently defined in the two references. 
 } \cite{Baumann:2007zm}
 \begin{align}
 \langle h_{\textbf{k}}(\eta)h_{\textbf{k}^{\prime}}(\eta') \rangle = \frac{16}{a^{2}(\eta)}\int^{\eta}_{\eta_{0}} d\tilde{\eta_{2}}\int^{\eta}_{\eta_{0}} d\tilde{\eta_{1}}~ a(\tilde{\eta_{1}})a(\tilde{\eta_{2}})\nonumber \\ \times ~G_{\textbf{k}}(\eta,\tilde{\eta_{1}})G_{\textbf{k}'}(\eta,\tilde{\eta_{2}})\langle S_{\textbf{k}}(\eta)S_{\textbf{k}'}(\eta') \rangle.
 \label{eq:hsrel}
 \end{align}
 Morever, to estimate the correlation function, $\langle S_{\textbf{k}}(\eta)S_{\textbf{k}'}(\eta') \rangle$, we do not consider non-Gaussanity in the primordial curvature power spectrum. With this assumption, we equate Eq. (\ref{eq:hpowdef}) with Eq. (\ref{eq:hsrel}) and after some 
 simplification obtain
 \cite{Baumann:2007zm,Kohri:2018awv}
 \begin{align}
 \mathcal{P}_{h}(\eta,k) = 4\int\limits_{0}^{\infty} dv \int\limits_{|1-v|}^{|1+v|} du \left[ \frac{4v^{2}-(1+v^{2}-u^{2})^{2}}{4vu}\right]^{2} \nonumber \\ \times  ~\mathrm{I}^{2}(v,u,x)P_{\zeta}(kv)P_{\zeta}(ku)
 \label{eq:phuv}
 \end{align} 
 where, $u=| \textbf{k}-\tilde{\textbf{k}}|/\mathrm{k}$ and $v=\tilde{\mathrm{k}}/\mathrm{k}$ are the dimensionless variable in which $\tilde{\mathrm{k}}$ corresponds for the wave vector associated with the scalar source $\Phi_{\tilde{\textbf{k}}}$.  In this expression, $x\equiv k\eta$ and the function $\mathrm{I}(v,u,x)$ is given as  
 \begin{equation}
 \mathrm{I}(v,u,x)=\int_{0}^{x} d\tilde{x}~ \frac{a(\tilde{\eta})}{a(\eta)} ~\mathrm{k} ~G_{\textbf{k}}(\eta,\tilde{\eta})~f(v,u,\tilde{x}).
 \end{equation}
 The function  $f(v,u,\tilde{x})$ consists of $\Phi$ terms and 
 is given in Ref. \cite{Kohri:2018awv}.
 Further, after redefining the variable $s=u-v$ and $t=u+v-1$, we may rewrite the Eq. (\ref{eq:phuv}) as
 \begin{align}
 \mathcal{P}_{h}(\eta,k) = & 2 \int\limits_{0}^{\infty} dt \int\limits_{-1}^{1} ds \left[ \frac{t(t+2)(s^{2}-1)}{(1+t+s)(1+t-s)}\right]^{2} \nonumber \\ & \times ~\mathrm{I}^{2}(v,u,x)P_{\zeta}(kv)P_{\zeta}(ku).
 \label{eq:phst}
 \end{align} 

 The gravitational wave energy density 
  defined as 
 $\rho_{\mathrm{GW}}(\eta)=\int d\ln \mathrm{k}~\rho_{\mathrm{GW}}(\eta,\mathrm{k})$
 can be estimated for the subhorizon modes as \cite{Maggiore:1999vm}
 \begin{equation}
 \rho_{\mathrm{GW}}=\frac{\mathrm{M^{2}_{pl}}}{16a^{2}}\langle \overline{h_{ij,k}~ h_{ij,k}} \rangle
 \end{equation}
 where the overline correspond to the  oscillation average. The  fraction of gravitational wave energy density per logarithmic $k$ interval to the total energy is given by
 \begin{align}
 \Omega_{\mathrm{GW}}(\eta,k)=&\frac{1}{\rho_{\mathrm{tot}}(\eta)}\left( \frac{d\rho_{\mathrm{GW}}(\eta)}{d\ln \mathrm{k}}\right) =\frac{\rho_{\mathrm{GW}}(\eta,\mathrm{k})}{\rho_{\mathrm{tot}}(\eta)}\nonumber \\=&\frac{1}{24}\left( \frac{\mathrm{k}}{a(\eta)H(\eta)}\right)^{2} \overline{P_h(\eta, \mathrm{k})}
 \label{eq:omegagw}
 \end{align}
 where $\rho_{\mathrm{tot}}(\eta)$ is the total energy density and $\overline{P_h(\eta, \mathrm{k})}$ is the  dimensionless power spectrum averaged over time, estimated using Eq. (\ref{eq:phst}). In this calculation, we have summed over both the polarization modes.
 For a  radiation dominated Universe, in the late time limit, $x\rightarrow \infty$, the function $\overline{\mathrm{I}^{2}(v,u,x)}$ can be simplified as \cite{Kohri:2018awv}
  \begin{align}
  &\overline{\mathrm{I}^{2}(v,u,x\rightarrow \infty)} =  \frac{9}{2x^{2}} \left(\frac{u^{2}+v^{2}-3}{4u^{3}v^{3}} \right)^{2} \times \nonumber \\ &\bigg[\left(-3uv+ (u^{2}+v^{2}-3)\log  \left| \frac{3-(u+v)^{2}}{3-(u-v)^{2}}\right|\right)^{2}    \bigg. \nonumber\\   \bigg.  &+  \pi^{2}(u^{2}+v^{2}-3)^{2}\theta(u+v-\sqrt{3}) 
  \bigg].
  \label{IRD}
  \end{align}
 Substituting Eq. (\ref{IRD}) in Eq. (\ref{eq:phst}) and carrying out the integral for $\mathcal{P}_{h}(\eta,k)$, we finally obtain $\Omega_{\mathrm{GW}}(\eta,k)$ from Eq. (\ref{eq:omegagw}).
  
 The observationally relevant quantity is the energy spectrum of induced gravitational waves $\Omega_{GW,0}(k)$ at the present time given by 
 \begin{equation}
 \Omega_{\mathrm{GW},0}(k)=0.39\left(\frac{g_{\star}(T_{c})}{106.75} \right)^{-\frac{1}{3}} \Omega_{r,0}~ \Omega_{\mathrm{GW}}(\eta_{c},k)
 \label{eq:omegagw0}
 \end{equation}
 where $\Omega_{r,0}h^{2} = 4.18\times 10^{-5} $
 is the present radiation energy density, and  $g_{\star}(T_{c}) $ is the effective number of
 relativistic degree of freedom in the radiation dominated era. Also, $\eta_{c}$ is the conformal time at the epoch when perturbation is inside the horizon after re-entry during radiation dominated era. Morever, the frequency of gravitational wave is related with the comoving scale as 
 \begin{equation}
 f=\frac{k}{2\pi}=1.5\times 10^{-15}\left( \frac{k}{1~ \rm{Mpc}^{-1}}\right)  \rm{Hz}.
 \label{eq:frkrel}
 \end{equation}
 Using this relation, we express $\Omega_{\mathrm{GW},0(k)}$ in terms of frequency of the gravitational wave.

 \section{Analysis and Discussion}
 \label{Sec4}
We consider a monomial potential ($V(\phi)=\lambda\phi^4$) of warm inflation with the dissipation coefficient $\Upsilon=C_\phi T^3/\phi^2$. Monomial potentials of inflation are single parameter models and are interesting as they predict a large value of primordial gravitational waves, which leads to their testability in future CMB experiments. In our earlier work \cite{Arya:2017zlb}, we had considered this model and estimated the parameter space of model variables consistent with the CMB observations. We found that unlike cold inflation, where $\lambda\phi^4$ potential is ruled out, there is a parameter space in warm inflation for which this potential can be a viable model for describing inflation. Also, the predicted value of the tensor-to-scalar ratio in this model can be tested in the future CMB polarization experiments.
 Next, in another work \cite{Arya:2019wck}, we found that interestingly this warm inflation model has features with a blue-tilted spectrum and a large amplitude of the primordial power spectrum at the small scales. This leads to the formation of PBHs with mass $M_{PBH}\sim 10^3$ g.  
 Further, we found that some parameter space of our warm inflation model is consistent with the bounds on the PBH mass fraction and thus interesting to explore. Furthermore, we expect that associated with the enhanced scalar power spectrum, there would be second order tensor modes.
  To complete the picture, we now extend our previous studies and explore the secondary induced gravitational waves from our warm inflation model.
 
 \subsection{Primordial Curvature Power Spectrum}
\begin{figure}[]
   \includegraphics[height=2.5in,width=3.1in]{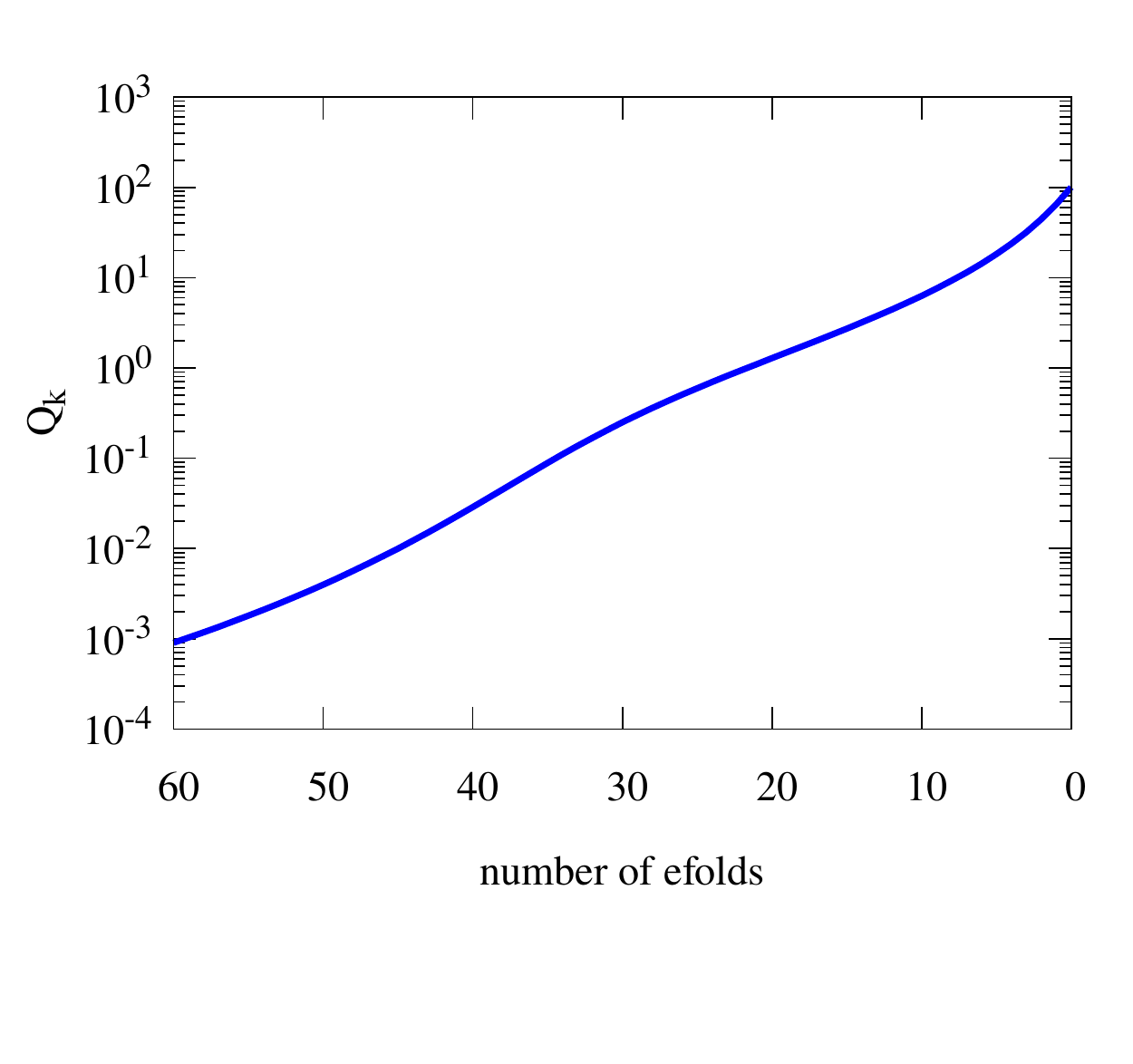}
   \includegraphics[height=2.in,width=2.9in]{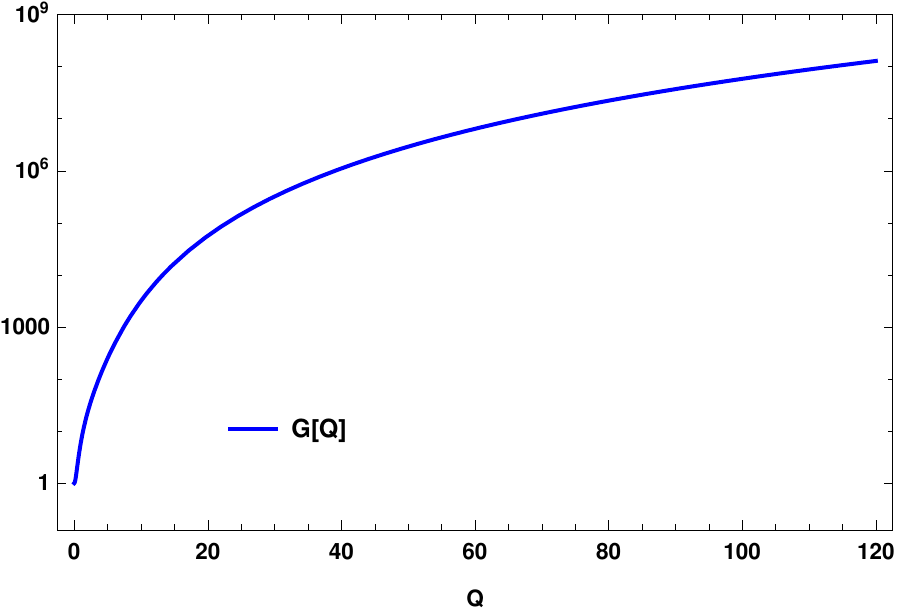} 
   \caption{\textit{Top: }The evolution of dissipation parameter $Q_k$ versus number of efolds of inflation for our warm inflation model and \textit{Bottom:} the growth factor $G(Q)$ versus $Q$, given in Eq. (\ref{GQ}) are shown here. }
   	\label{gq}	 
   \end{figure}

\begin{figure}[]
 	\centering
 \includegraphics[height=2.in,width=3.in]{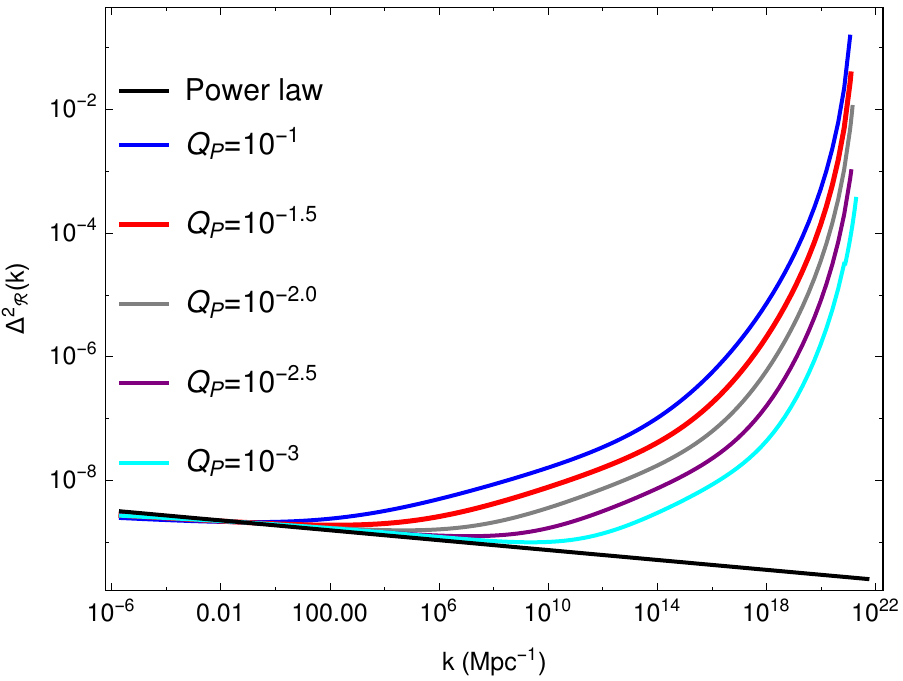}
 \caption{The primordial curvature power spectrum for different values of dissipation parameter $Q_P$ as a function of the scale $k$. Figure taken from Ref. \cite{Arya:2019wck}. }
 	\label{fig:pk}	
 \end{figure}

 We first show the evolution of dissipation parameter as a function of number of efolds of inflation and the growth factor $G(Q_k)$ in Fig. \ref{gq}. Here the number of efolds are counted from the end of inflation ($N_e=0$) such that the pivot scale corresponds to $N_P=60$. 
 In this model of warm inflation, we have \cite{Arya:2017zlb}
 \begin{equation} 
 	\frac{dQ}{dN}= -40 \left(\frac{9(\pi^2 g_*/30)^3}{64C_{\phi}^4\lambda}\right)^\frac{1}{5}\frac{Q^{6/5}(1+Q)^{6/5}}{(1+7Q)}.
 	\label{dQdN1}
 	\end{equation}
 	The negative sign of $dQ/dN$ implies that the dissipation parameter $Q$ increases as the inflation proceeds such that it evolves from weak dissipative regime at the pivot scale to a strong dissipative regime near the end of inflation, as can be seen in Fig. \ref{gq}. As the growth function is proportional to the dissipation parameter through Eq. (\ref{GQ}), there is a huge enhancement in the $G(Q)$ (Fig. \ref{gq}) and subsequently the primordial power spectrum near the end of inflation, as shown in Fig. \ref{fig:pk}.
 
 We can see from Fig. \ref{fig:pk} that the primordial power spectrum is red-tilted ($n_s<1$) for the CMB scales and for some range of $Q_P$ values, it is consistent with the $n_s-r$ bounds from Planck observations.
 Alongwith, it has features that it is blue-tilted ($n_s>1$) at the small scales with a large amplitude, due to a large growth factor. The enhanced amplitude of scalar fluctuations at small scales then source the formation of primordial black holes and furthermore secondary gravitational waves.
 
 \subsection{Spectral index and tensor-to-scalar ratio}
\begin{figure}[]
 \includegraphics[height=2.in,width=3.in]{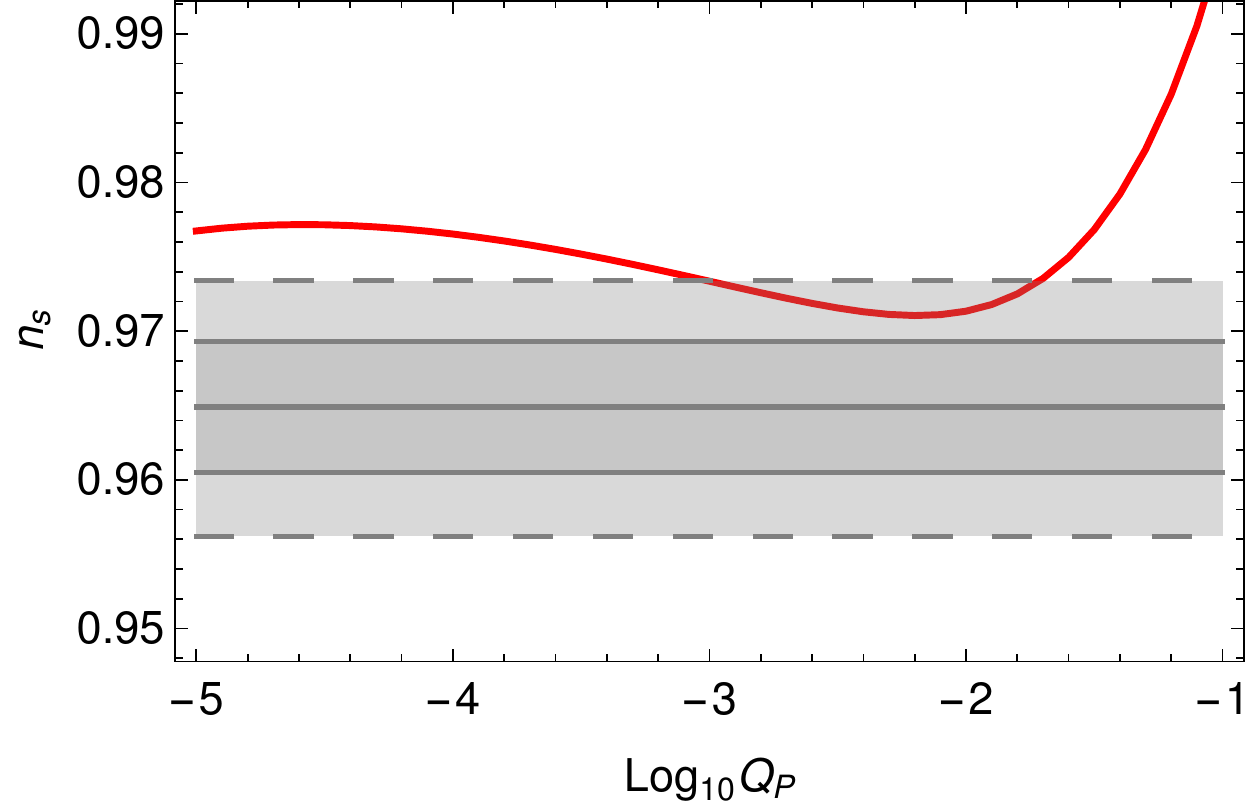}  
 \includegraphics[height=2.in,width=3.in]{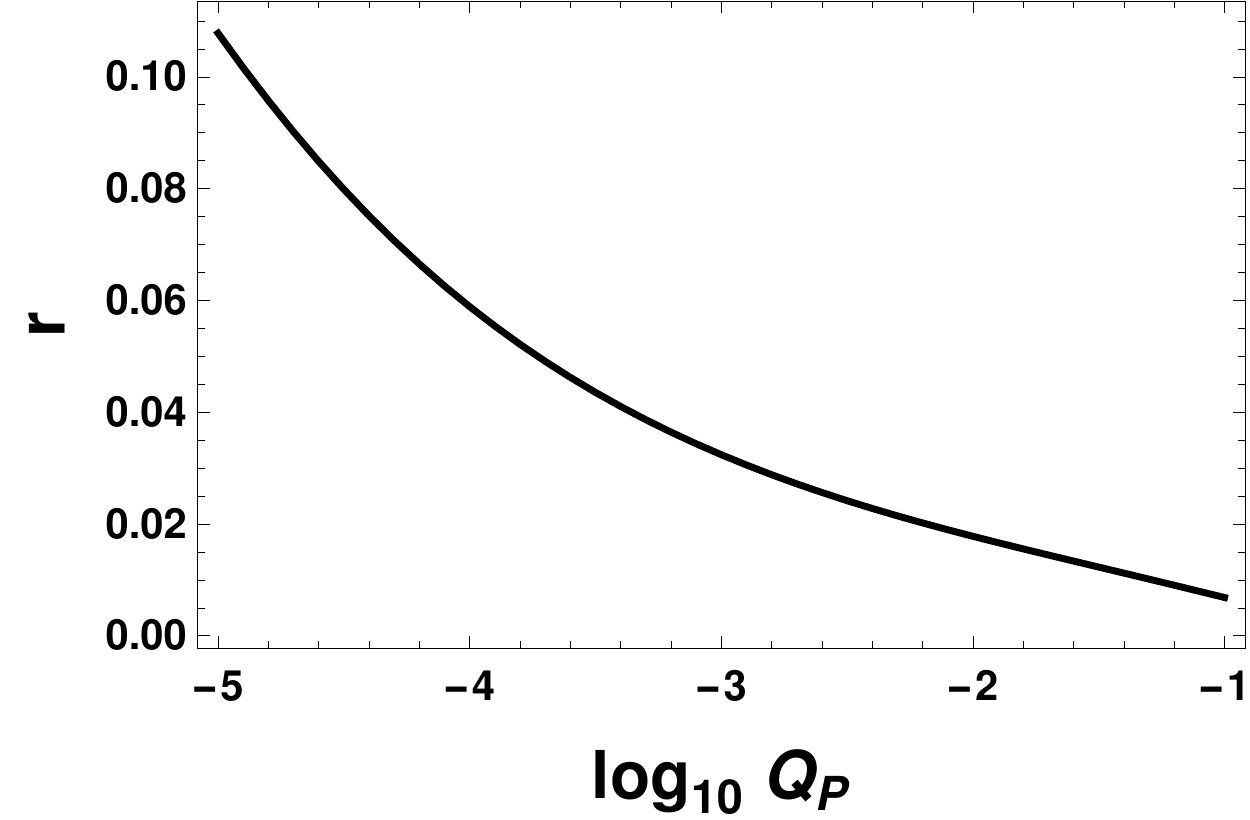} 
 \caption{\textit{Top:} The spectral index or tilt of the primordial curvature power spectrum for our warm inflation model.  The colored band represents the allowed $1-2 \sigma$ range of $n_s$ from the Planck observations. Here the number of efolds of inflation equals to 60.  
 \textit{Bottom:}  Tensor-to-scalar ratio as a function of $Q_P$ is plotted here.}
 	\label{fig:ns}	
 \end{figure}

 In Fig. \ref{fig:ns}, we plot the scalar spectral index, defined as the tilt of primordial power spectrum at the pivot scale, 
 \begin{equation}
 n_s-1\equiv\left. \frac{d\ln\Delta^2_\mathcal{R}(k)}{d\ln k}\right\rvert_{k=k_P}.
 \end{equation}
 We see that only weak dissipative regime can be consistent with the $n_s$ values in this model.
 In the same Figure, we also plot the tensor-to-scalar ratio, defined as the ratio of the amplitude of the tensor power spectrum to the amplitude of the scalar power spectrum at the pivot scale
 \begin{equation}
 r\equiv \frac{\Delta^2_t(k_P)}{\Delta^2_\mathcal{R}(k_P)}.
 \end{equation}
 We can see that the value of $r$ decreases as the dissipation parameter increases.

We find that for parameter space $[6.31 \times 10^{-4}<Q_P<0.02]$, our warm inflation model is consistent with the observationally allowed $n_s-r$ values.  Therefore, we will explore the small scale features of our warm inflation model for this range of $Q_P$ values.
 
 \vspace{-0.2cm}
 \subsection{Induced Gravitational Wave Spectrum}
\begin{figure}[]
  		\centering
  		\includegraphics[height=2.2in,width=3.1in]{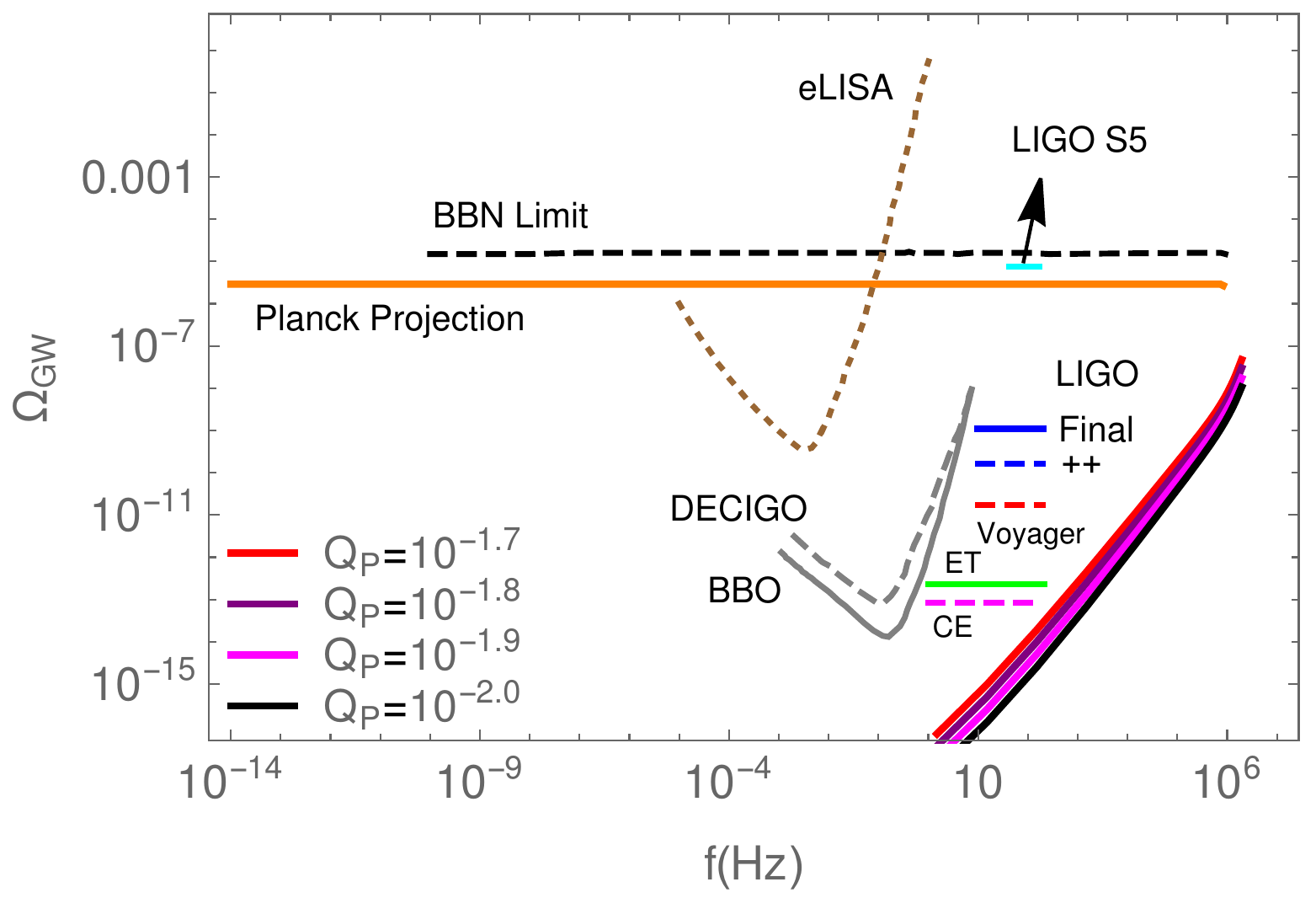} 
  \caption{The spectrum of secondary gravitatonal waves $\Omega_{GW}$ at the present time, generated for our warm inflation model as a function of the frequency of gravitational wave.  The sensitivity plots of various detectors are taken from Ref. \cite{Mandic} and references therein.}	
  	\label{fig:gw}
  	\end{figure}

 Using Eqs. (\ref{eq:phst}), (\ref{eq:omegagw}), and (\ref{eq:omegagw0}), we calculate the induced gravitational wave spectrum for the primordial power spectrum of warm inflation given in Eq. (\ref{power}).
 Then using Eq. (\ref{eq:frkrel}),  we plot the present spectral energy density of the produced secondary gravitational waves as a function of the frequency in Fig. \ref{fig:gw}. We also include the theoretical constraints on $\Omega_{GW,0}$ and sensitivity curves for the present and future gravitational wave detectors in the figure. For details, see Ref. \cite{Maggiore:1999vm,Assadullahi:2009jc,2009Natur.460..990A,Moore:2014lga,Inomata:2018epa,Mandic} and references therein. 
 Here is a summary of various constraints on the spectral energy density and sensitivities of different gravitational wave detectors. 
 \subsubsection*{Constraints on the Primordial Gravitational Wave Background}
 \begin{itemize}
 \item \textbf{Big-Bang Nucleosynthesis (BBN)}: The presence of large amplitude of gravitational waves at the time of BBN, alters the light nuclei abundances predicted by the standard BBN. 
 This gives a constraint on the GW background as,
 $\Omega_{\mathrm{GW},0}<1.5\times 10^{-5}$ corresponding to frequency $\nu>10^{-10}$ Hz today \cite{Maggiore:1999vm}\cite{LIGOScientific:2006zmq}.
 \item \textbf{Cosmic Microwave Background (CMB): } The presence of extra relativistic degrees of freedom or large energy density of GW at the time of recombination may change the epoch of matter-radiation equality and  acoustic oscillations \cite{Sendra:2012wh}. This gives an upper bound on the GW background as 
 $\Omega_{\mathrm{GW},0}<2.7\times 10^{-6}$ corresponding to frequencies $\nu> 10^{-15}$ Hz \cite{Smith:2006nka,Pagano:2015hma}.
 \item \textbf{Second generation GW detectors - LIGO/VIRGO  \cite{KAGRA:2013rdx}: }
 The bounds on energy density of the gravitational waves from direct ground based detectors are  $\Omega_{\mathrm{GW},0}<6.9\times 10^{-6}$  through LIGO S$5$ run with a  maximum sensitivity at frequency $\sim 100$ Hz \cite{2009Natur.460..990A}, $\Omega_{\mathrm{GW},0}< 10^{-9}$ in Advanced LIGO/VIRGO  at frequency $\nu\sim30$ Hz \cite{Parikh:2020fhy}, $\Omega_{\mathrm{GW},0}< 10^{-10}$ in LIGO$++$ \cite{Ligoandvoyager:2022}.
 \item \textbf{Space-based GW detectors - LISA \cite{2017arXiv170200786A}, BBO, DECIGO \cite{Yagi:2011wg}: }
 LISA is a planned space GW detector, which is expected to detect primordial gravitational wave background to 
 $\Omega_{\mathrm{GW},0}<10^{-10}$ at $\nu\sim 1$ mHz \cite{Hogan:2001jn}\cite{Cornish:2001bb}.
 BBO \cite{Corbin:2005ny} and DECIGO \cite{Sato:2017dkf} are also future projects with proposed ability to detect GW down to $\Omega_{\mathrm{GW},0}\approx 10^{-16}$ at frequency $\nu\sim 1$ Hz.
 \item \textbf{Pulsar Timing Array (PTA) Experiments \cite{Lentati:2015qwp}: } Pulsars can be used as very stable clocks. These objects emit radio signals as pulses. By measuring the time of arrival of the radio pulses on Earth, one can detect perturbations due to gravitational waves. For $\nu\sim 10^{-8}$ Hz, the bound on GW energy density is $\Omega_{\mathrm{GW},0}< 4\times 10^{-8}$ \cite{Jenet:2006sv}.
 \item \textbf{Third generation ground-based GW detectors - Einstein Telescope \cite{ET}\, Cosmic Explorer \cite{2015PhRvD..91h2001D}:   }
 ET is a proposed  project with three detectors in triangle geometry, similar to LISA, whereas CE is a L-shaped geometry similar to advanced LIGO. The ET puts a limit on GW energy density $\Omega_{\mathrm{GW},0}<4\times  10^{-13}$ and from CE the bound is $\Omega_{\mathrm{GW},0}<  1.6\times 10^{-13}$.
 \end{itemize}
 
 In Fig. \ref{fig:gw}, the solid colored lines (red, blue, magenta, black) correspond to various values of the parameter $Q_P$ for our warm inflation model ($Q_P=10^{-1.7}, 10^{-1.8}, 10^{-1.9}, 10^{-2.0}$, respectively).
  For these $Q_P$ values, our model simultaneously explain the large scale CMB observations, as well as leads to the  generation of PBHs and SIGW at the small scales.
  From the figure,  it is evident that for stronger dissipation (large $Q_P$), the strength of the gravitational waves $\Omega_{GW}$ is large. Also, it is important to note that a majority of contribution to the primordial gravitational wave spectrum in Eq. (\ref{eq:phst}) 
 comes from the modes exiting the horizon near the end of inflation. These small scale modes with a large amplitude 
 induce significant secondary gravitational waves over a frequency range $f=(1-10^{6})$ Hz. This behavior can also be confirmed from Eq. (\ref{eq:frkrel}). 
 
 Further, from the sensitivity curves of various observatories plotted in Fig. \ref{fig:gw}, we infer that 
 the scalar induced gravitation waves produced from our model is quite feeble to be detected withing the current and near future 
 interferometer GW detectors, operational for frequencies typically less than $\sim10$ kHz. However, with the new detection techniques designed for comparably larger frequency range, 
 such as levitated-sensor-based gravitational-wave detector \cite{Aggarwal:2020umq}, microwave cavities \cite{Bernard:2002ci}, decameter Michelson interferometers \cite{Holometer:2016qoh}, resonant mass detectors {\cite{Aguiar:2010kn}} (also see Refs. \cite{Aggarwal:2020olq,Berlin:2021txa} for a comprehensive review), 
 one hopes to scrutinize these models more efficiently and better understand the rich physics of the early Universe.
  
 \section{Summary}
 \label{Sec5}
 The inflationary paradigm of early universe uniquely predicts a spectrum of primordial gravitational waves. These have not been detected yet, however are important to understand the inflationary physics. Likewise, the small scale spectrum of primordial perturbations is not well measured and therefore an important aspect to explore the inflationary dynamics. Primordial black holes are one such remarkable probe of the small scales, that provide constrains on the primordial curvature power spectrum, and thus different inflationary models. 
 As the amplitude of primordial curvature power spectrum is enhanced by many orders of magnitude for the formation 
 of PBHs, there are also associated second-order tensor fluctuations sourced by the scalar fluctuations. The focus of this paper is to study these scalar induced gravitational waves 
 from a model of warm inflation. 
 
 Warm Inflation is a well-motivated  and general description of inflation where the dissipative and non-equilibrium processes are present during the inflationary phase.
 In this scenario, the inflaton dissipates into radiation fields during inflation, which modifies both the background inflaton dynamics as well as its perturbations. 
The primordial power spectrum of warm inflation is sourced dominantly by the thermal fluctuations, and thus has different predictions of observables than the cold inflation. 
Here we discuss a model of warm inflation with a quartic potential and a dissipation coefficient $\Upsilon\propto T^3$. 
In this model, we find that the dissipation parameter increases from weak dissipative regime at the pivot scale ($Q<1$) to strong dissipative regime near the end of inflation ($Q\gg 1$). Accordingly, the growth factor in the primordial power spectrum $G(Q)$, characterizing the backreaction of radiation fluctuations to the inflaton fluctuations, also increase tremendously.
Thus,
for certain parameter space of this model, there is a huge growth in the scalar curvature power spectrum on very small scales, leading to the formation of primodial black holes, as well as secondary gravitational waves.

We find that the strength of these secondary gravitational waves is directly proportional to the dissipation parameter, i.e. a large amplitude of present GW energy density implies a stronger dissipation, and vice-versa. Also, the modes exiting the horizon near the end of inflation with a large amplitude contribute majorly to the GW energy density. This corresponds to a GW spectrum over the frequency range $1-10^6$ Hz in our model. 
It is found that the generated spectrum does not lie in the sensitivities of different ongoing and future laser interferometer GW detectors. However, some more sensitive GW detectors, such as the levitated-sensor detector, microwave cavities, decameter Michelson interferometers, resonant mass detectors, will explore the high frequency spectrum of our proposed model and might test the feasability of warm inflation model in future.  

\vspace{-1cm}
\section{Acknowledgement}
We would like to thank Prof. Raghavan Rangarajan, Prof. Namit Mahajan, and Prof. Rajeev Jain for useful discussions and suggestions. We also thank Physical Research Laboratory, Ahmedabad, India, for giving us the platform to carry out this research. Work of RA is supported by the National Post-Doctoral Fellowship by SERB, Government of India. AKM acknowledges the support
through Ramanujan Fellowship (PI: Dr. Diptimoy Ghosh) offered by the Department of Science and Technology, Government of India.
\appendix
\section{Calculation of dissipation coefficient for our warm inflation model}
\label{A}
Here we follow the lecture notes \cite{lec} and Refs. \cite{ Moss:2006gt,Berera:2008ar,BasteroGil:2010pb,BasteroGil:2012cm} to show the calculation of dissipation coefficient for our model. We also refer the reader to see Refs. \cite{Berera:1998gx,Bastero-Gil:2018yen,Das:2020xmh,Bastero-Gil:2021fac,supp,BASTEROGIL2021136055} for derivation of dissipation coefficient in other warm inflation models.

On accounting for the interactions of the inflaton with intermediate scalar boson $\chi$ and fermion $\psi_\chi$, as given in Eqs. (\ref{scalar}), (\ref{yukawa}), the dissipation coefficient at leading order is given as \cite{BasteroGil:2010pb}
 \begin{align}
 \Upsilon=\frac{2}{T} g^4 \phi^2 \int \frac{d^4p}{(2\pi)^4}~ [\rho_{\chi_1} (\omega ,\boldsymbol{p})^2+\rho_{\chi_2} (\omega ,\boldsymbol{p})^2]~\nonumber \\ \hspace{1cm} \times n_B(\omega)~(1+n_B(\omega)) \nonumber \\
 +\frac{2}{T} g^2\int \frac{d^4p}{(2\pi)^4}~ tr[\rho_{\psi_\chi}(\omega ,\boldsymbol{p})^2]~n_F(\omega)~(1-n_F(\omega)).
 \label{upsilo}
 \end{align}
 Here $n_B(\omega)$, $n_F(\omega)$ are the Bose-Einstein and Fermi-Dirac distributions, respectively, and $\rho_\chi$, $\rho_{\psi_\chi}$ are the spectral functions for the intermediate $\chi$, $\psi_\chi$ fields.

 \begin{align}
 \rho_\chi(\omega,\boldsymbol{p})~=&~\frac{i}{p^2+m^2_{\chi,R}+{i\rm{Im}\Sigma_{\chi}}}-\frac{i}{p^2+m^2_{\chi,R}-{i\rm{Im}\Sigma_{\chi}}}\nonumber \\ 
 \vspace{1.5cm}
 ~=&~\frac{2{\rm{~Im}\Sigma_{\chi}} }{(p^2+m^2_{\chi,R})^2+({\rm{Im}\Sigma_{\chi}})^2}\nonumber \\ 
 \vspace{1.5cm}
 =& \frac{4 ~\omega_p\Gamma_\chi}{(-\omega^2+\omega_p^2)^2+4~\omega_p^2~\Gamma_\chi^2},
 \label{spectralb}
 \end{align}
 where $\Gamma_\chi$ is the decay width of the $\chi$ field and is related to the imaginary component of the self energy $\Sigma_\chi$,  $\omega_p^2=|\boldsymbol{p}^2|+m^2_{\chi,R}$ is the dispersion relation of the $\chi$ field, and $m^2_{\chi,R}=m_\chi^2+\rm{Re}\Sigma_\chi$ is the effective, renormalized mass of the $\chi$ field.
 
 The spectral function for fermionic field $\psi_\chi$ is given by
 \begin{align}
 \rho_{\psi_\chi}(\omega,\boldsymbol{p})=\frac{i}{\slashed p+m_{{\psi_\chi},R}+i \rm{Im}\Sigma_{\psi_\chi} }-\frac{i}{\slashed p+m_{{\psi_\chi},R}-i \rm{Im}\Sigma_{\psi_\chi} },
  \end{align}
  where $m_{\psi_\chi,R}=m_{\psi_\chi}+\rm{Re}\Sigma_{\psi_\chi}$ is the effective, renormalized mass, and $\Sigma_{\psi_\chi}$ is the self energy of the $\psi_\chi$ field.

  Thus, to calculate the dissipation coefficient, we need to compute the masses of $\chi,\psi_\chi$ fields and their decay width at finite temperature. 
  The decay width of the $\chi,\psi_\chi$ fields has contributions from direct, inverse as well as thermal scatterings (Landau damping). The response timescale of the system is associated with the decay width as $\tau\rightarrow 1/\Gamma.$
 For explicit calculations and expressions of the field self energy and decay widths, see Ref. \cite{BasteroGil:2010pb}.

 In certain approximations, the dissipation coefficient given in Eq. (\ref{upsilo}) reduces to simplified expression.
 In one regime, the poles of the spectral function dominate the integral and is called the pole approximation. In the other regime, the integration is limited to low-momentum and it is referred to as the low-momentum approximation. 
 \subsubsection{Low temperature limit}
 	In this regime, the temperature of the thermal bath is much less than the masses of the intermediate catalyst fields, $\chi$ and $\psi_\chi$, i.e. 
 	$T\ll m_{\chi,R},m_{\psi_\chi,R}$, but is higher compared to the radiation fields, $T\gg m_{\sigma,R},m_{\psi_\sigma,R}$. The thermal corrections to the effective masses of the $\chi,\psi_\chi$ fields can be neglected in this regime, i.e. $m_{\chi,R}^2\simeq m_{\chi}^2=2g^2\varphi^2$ and 
 	$m_{\psi_\chi,R}^2\simeq m_{\psi_\chi}^2=2g^2\varphi^2.$
 For large values of $m_{\chi}/T,$ the dominant contributions to the dissipation coefficient given in Eq. (\ref{upsilo}) come from virtual $\chi$ fields with low energy and momentum, $\omega$, $|\boldsymbol{p}|\sim T\ll m_{\chi}$ which leads to the low-momentum approximation.
 	Then, $(\omega^2-\omega_p^2)^2\approx m_{\chi}^4$, and the spectral function for the scalar boson in Eq. (\ref{spectralb}) becomes
 	$\rho_\chi \simeq \frac{4}{m_{\chi}^3}\Gamma_\chi,$ 
 	which gives a leading order contribution to the dissipation coefficient, given in Eq. (\ref{upsilo}), $\propto T^3/m_{\chi}^2$ \cite{Moss:2006gt,BasteroGil:2010pb,BasteroGil:2012cm}.
 	The fermionic contribution is calculated to be subleading in $T$ in the low temperature limit ($\propto T^5/m_{\psi_\chi}^4$) \cite{Moss:2006gt,BasteroGil:2010pb}.
                            	
 A detailed analysis including thermal corrections to the $\chi$ mass and finite decay width of $\chi$ field in the spectral function gives \cite{BasteroGil:2012cm}	\vspace{-0.3cm}
 \begin{equation}
 	\Upsilon~=~C_\phi \frac {T^3}{\phi^2},
 	\label{upsiloncubic}
 		\vspace{-0.1cm}
 	\end{equation}
	where $C_\phi= \frac{h^2 }{16 \pi} N_Y N_X$, which depends on the multiplicities of $X$ and $Y$ superfields and coupling between them. 
	\vspace{1cm}
\subsubsection{High temperature limit}
 	In this limit, the intermediate catalyst fields are lighter, $ m_{\chi,R},m_{\psi_\chi,R}\ll T$. The main contribution to the dissipation coefficient comes from the pole in the spectral function at $\omega=\omega_p$ and a resonant production of on-shell $\chi$ particles take place.
 	In the pole approximation, the bosonic spectral function becomes
 	$\rho_\chi^2\rightarrow\frac{\pi}{2~\omega_p^2\Gamma_\chi}\delta(\omega-\omega_p).$ 
 	Substituting this in Eq. (\ref{upsilo}) for the scalar field, the dissipation coefficient gets a contribution which is linearly dependent on the temperature of the thermal bath $\Upsilon\approx0.691~ \frac{g^2}{h^2} T$ \cite{Moss:2006gt}.
 	On accounting all the fermionic and bosonic contributions in Eq. (\ref{upsilo}), the total dissipation coefficient is obtained to be \cite{Moss:2006gt} \vspace{-0.3cm}
 	\begin{equation}
 	\Upsilon~=~C_T T, \hspace{2cm} C_T\approx0.97~ \frac{g^2}{h^2}.
 	\label{linearupsilon}\vspace{-0.3cm}
 	\end{equation}
 	By knowing the value of $C_T$, one can calculate the order of ratio of couplings $g/h$, which is useful in model building. 
 	
\bibliographystyle{utphys}
\bibliography{SIGW_WI_v1}

\providecommand{\href}[2]{#2}\begingroup\raggedright\begin{thebibliography}{100}

\bibitem{Arya:2019wck}
R.~Arya, ``{Formation of Primordial Black Holes from Warm Inflation},''
  \href{http://dx.doi.org/10.1088/1475-7516/2020/09/042}{{\em JCAP} {\bfseries
  09} (2020) 042}, \href{http://arxiv.org/abs/1910.05238}{{\ttfamily
  arXiv:1910.05238 [astro-ph.CO]}}.

\bibitem{Kazanas:1980tx}
D.~Kazanas, ``{Dynamics of the Universe and Spontaneous Symmetry Breaking},''
\href{http://dx.doi.org/10.1086/183361}{{\em Astrophys. J.} {\bfseries 241}
  (1980) L59--L63}.

\bibitem{Sato:1980yn}
K.~Sato, ``{First Order Phase Transition of a Vacuum and Expansion of the
  Universe},''
{\em Mon. Not. Roy. Astron. Soc.} {\bfseries 195} (1981) 467--479.

\bibitem{Guth:1980zm}
A.~H. Guth, ``{The Inflationary Universe: A Possible Solution to the Horizon
  and Flatness Problems},''
  \href{http://dx.doi.org/10.1103/PhysRevD.23.347}{{\em Phys. Rev.} {\bfseries
  D23} (1981) 347--356}.
[Adv. Ser. Astrophys. Cosmol.3,139(1987)].

\bibitem{Starobinsky:1980te}
A.~A. Starobinsky, ``{A New Type of Isotropic Cosmological Models Without
  Singularity},'' \href{http://dx.doi.org/10.1016/0370-2693(80)90670-X}{{\em
  Phys. Lett.} {\bfseries 91B} (1980) 99--102}.
[,771(1980)].

\bibitem{Linde:1981mu}
A.~D. Linde, ``{A New Inflationary Universe Scenario: A Possible Solution of
  the Horizon, Flatness, Homogeneity, Isotropy and Primordial Monopole
  Problems},'' \href{http://dx.doi.org/10.1016/0370-2693(82)91219-9}{{\em Phys.
  Lett.} {\bfseries 108B} (1982) 389--393}.
[Adv. Ser. Astrophys. Cosmol.3,149(1987)].

\bibitem{Akrami:2018odb}
{\bfseries Planck} Collaboration, Y.~Akrami {\em et~al.}, ``{Planck 2018
  results. X. Constraints on inflation},''. arXiv:1807.06211.

\bibitem{BICEP:2021xfz}
{\bfseries BICEP, Keck} Collaboration, P.~A.~R. Ade {\em et~al.}, ``{Improved
  Constraints on Primordial Gravitational Waves using Planck, WMAP, and
  BICEP/Keck Observations through the 2018 Observing Season},''
  \href{http://dx.doi.org/10.1103/PhysRevLett.127.151301}{{\em Phys. Rev.
  Lett.} {\bfseries 127} no.~15, (2021) 151301},
  \href{http://arxiv.org/abs/2110.00483}{{\ttfamily arXiv:2110.00483
  [astro-ph.CO]}}.

\bibitem{Baumann:2009ds}
D.~Baumann,
  \href{http://dx.doi.org/10.1142/9789814327183_0010}{``{Inflation},''} in {\em
  {Physics of the large and the small, TASI 09, 1-26 June 2009}}, pp.~523--686.
\newblock 2011.
\newblock
\href{http://arxiv.org/abs/0907.5424}{{\ttfamily arXiv:0907.5424 [hep-th]}}.
\newblock

\bibitem{Linde:2007fr}
A.~D. Linde, ``{Inflationary Cosmology},''
  \href{http://dx.doi.org/10.1007/978-3-540-74353-8\_1}{{\em Lect. Notes Phys.}
  {\bfseries 738} (2008) 1--54},
  \href{http://arxiv.org/abs/0705.0164}{{\ttfamily arXiv:0705.0164 [hep-th]}}.

\bibitem{Tsujikawa:2003jp}
S.~Tsujikawa, ``{Introductory review of cosmic inflation},'' in {\em {2nd Tah
  Poe School on Cosmology}: {Modern Cosmology}}.
\newblock 4, 2003.
\newblock \href{http://arxiv.org/abs/hep-ph/0304257}{{\ttfamily
  arXiv:hep-ph/0304257}}.

\bibitem{Olive:1989nu}
K.~A. Olive, ``{Inflation},''
  \href{http://dx.doi.org/10.1016/0370-1573(90)90144-Q}{{\em Phys. Rept.}
  {\bfseries 190} (1990) 307--403}.

\bibitem{Riotto:2002yw}
A.~Riotto, ``{Inflation and the theory of cosmological perturbations},'' {\em
  ICTP Lect. Notes Ser.} {\bfseries 14} (2003) 317--413,
\href{http://arxiv.org/abs/hep-ph/0210162}{{\ttfamily arXiv:hep-ph/0210162
  [hep-ph]}}.

\bibitem{Mollerach:2003nq}
S.~Mollerach, D.~Harari, and S.~Matarrese, ``{CMB polarization from secondary
  vector and tensor modes},''
  \href{http://dx.doi.org/10.1103/PhysRevD.69.063002}{{\em Phys. Rev. D}
  {\bfseries 69} (2004) 063002},
  \href{http://arxiv.org/abs/astro-ph/0310711}{{\ttfamily
  arXiv:astro-ph/0310711}}.

\bibitem{Ananda:2006af}
K.~N. Ananda, C.~Clarkson, and D.~Wands, ``{The Cosmological gravitational wave
  background from primordial density perturbations},''
  \href{http://dx.doi.org/10.1103/PhysRevD.75.123518}{{\em Phys. Rev. D}
  {\bfseries 75} (2007) 123518},
  \href{http://arxiv.org/abs/gr-qc/0612013}{{\ttfamily arXiv:gr-qc/0612013}}.

\bibitem{Baumann:2007zm}
D.~Baumann, P.~J. Steinhardt, K.~Takahashi, and K.~Ichiki, ``{Gravitational
  Wave Spectrum Induced by Primordial Scalar Perturbations},''
  \href{http://dx.doi.org/10.1103/PhysRevD.76.084019}{{\em Phys. Rev. D}
  {\bfseries 76} (2007) 084019},
  \href{http://arxiv.org/abs/hep-th/0703290}{{\ttfamily arXiv:hep-th/0703290}}.

\bibitem{Saito:2008jc}
R.~Saito and J.~Yokoyama, ``{Gravitational wave background as a probe of the
  primordial black hole abundance},''
  \href{http://dx.doi.org/10.1103/PhysRevLett.102.161101}{{\em Phys. Rev.
  Lett.} {\bfseries 102} (2009) 161101},
  \href{http://arxiv.org/abs/0812.4339}{{\ttfamily arXiv:0812.4339
  [astro-ph]}}. [Erratum: Phys.Rev.Lett. 107, 069901 (2011)].

\bibitem{2010PThPh.123..867S}
R.~{Saito} and J.~{Yokoyama}, ``{Gravitational-Wave Constraints on the
  Abundance of Primordial Black Holes},''
  \href{http://dx.doi.org/10.1143/PTP.123.867}{{\em Progress of Theoretical
  Physics} {\bfseries 123} no.~5, (May, 2010) 867--886},
  \href{http://arxiv.org/abs/0912.5317}{{\ttfamily arXiv:0912.5317
  [astro-ph.CO]}}.

\bibitem{zeldo:1966}
Y.~B. Zel'dovich and I.~D. Novikov, ``{The Hypothesis of Cores Retarded During
  Expansion and the Hot Cosmological Model},'' {\em Soviet Astronomy}
  {\bfseries 10} (1967) 602.

\bibitem{Hawking:1971ei}
S.~Hawking, ``{Gravitationally collapsed objects of very low mass},''
{\em Mon. Not. Roy. Astron. Soc.} {\bfseries 152} (1971) 75.

\bibitem{Carr:1974nx}
B.~J. Carr and S.~W. Hawking, ``{Black holes in the early Universe},''
{\em Mon. Not. Roy. Astron. Soc.} {\bfseries 168} (1974) 399--415.

\bibitem{Carr:1975qj}
B.~J. Carr, ``{The Primordial black hole mass spectrum},''
\href{http://dx.doi.org/10.1086/153853}{{\em Astrophys. J.} {\bfseries 201}
  (1975) 1--19}.

\bibitem{Hawking:1982ga}
S.~W. Hawking, I.~G. Moss, and J.~M. Stewart, ``{Bubble Collisions in the Very
  Early Universe},''
\href{http://dx.doi.org/10.1103/PhysRevD.26.2681}{{\em Phys. Rev.} {\bfseries
  D26} (1982) 2681}.

\bibitem{Hogan:1984zb}
C.~J. Hogan, ``{Massive Black Holes generated by Cosmic Strings},''
\href{http://dx.doi.org/10.1016/0370-2693(84)90810-4}{{\em Phys. Lett.}
  {\bfseries 143B} (1984) 87--91}.

\bibitem{Hawking:1987bn}
S.~W. Hawking, ``{Black Holes From Cosmic Strings},''
\href{http://dx.doi.org/10.1016/0370-2693(89)90206-2}{{\em Phys. Lett.}
  {\bfseries B231} (1989) 237--239}.

\bibitem{Caldwell:1996pt}
R.~R. Caldwell, A.~Chamblin, and G.~W. Gibbons, ``{Pair creation of black holes
  by domain walls},'' \href{http://dx.doi.org/10.1103/PhysRevD.53.7103}{{\em
  Phys. Rev.} {\bfseries D53} (1996) 7103--7114},
\href{http://arxiv.org/abs/hep-th/9602126}{{\ttfamily arXiv:hep-th/9602126
  [hep-th]}}.

\bibitem{2010PhRvD..81b3517B}
E.~{Bugaev} and P.~{Klimai}, ``{Induced gravitational wave background and
  primordial black holes},''
  \href{http://dx.doi.org/10.1103/PhysRevD.81.023517}{{\em \prd} {\bfseries 81}
  no.~2, (Jan., 2010) 023517}, \href{http://arxiv.org/abs/0908.0664}{{\ttfamily
  arXiv:0908.0664 [astro-ph.CO]}}.

\bibitem{Alabidi:2012ex}
L.~Alabidi, K.~Kohri, M.~Sasaki, and Y.~Sendouda, ``{Observable Spectra of
  Induced Gravitational Waves from Inflation},''
  \href{http://dx.doi.org/10.1088/1475-7516/2012/09/017}{{\em JCAP} {\bfseries
  09} (2012) 017}, \href{http://arxiv.org/abs/1203.4663}{{\ttfamily
  arXiv:1203.4663 [astro-ph.CO]}}.

\bibitem{Inomata:2018cht}
K.~Inomata, M.~Kawasaki, K.~Mukaida, and T.~T. Yanagida, ``{Double inflation as
  a single origin of primordial black holes for all dark matter and LIGO
  observations},'' \href{http://dx.doi.org/10.1103/PhysRevD.97.043514}{{\em
  Phys. Rev. D} {\bfseries 97} no.~4, (2018) 043514},
  \href{http://arxiv.org/abs/1711.06129}{{\ttfamily arXiv:1711.06129
  [astro-ph.CO]}}.

\bibitem{Orlofsky:2016vbd}
N.~Orlofsky, A.~Pierce, and J.~D. Wells, ``{Inflationary theory and pulsar
  timing investigations of primordial black holes and gravitational waves},''
  \href{http://dx.doi.org/10.1103/PhysRevD.95.063518}{{\em Phys. Rev. D}
  {\bfseries 95} no.~6, (2017) 063518},
  \href{http://arxiv.org/abs/1612.05279}{{\ttfamily arXiv:1612.05279
  [astro-ph.CO]}}.

\bibitem{Ahmed:2021ucx}
W.~Ahmed, M.~Junaid, and U.~Zubair, ``{Primordial Black Holes and Gravitational
  Waves in Hybrid Inflation with Chaotic Potentials},''
  \href{http://arxiv.org/abs/2109.14838}{{\ttfamily arXiv:2109.14838
  [astro-ph.CO]}}.

\bibitem{Fu:2019vqc}
C.~Fu, P.~Wu, and H.~Yu, ``{Scalar induced gravitational waves in inflation
  with gravitationally enhanced friction},''
  \href{http://dx.doi.org/10.1103/PhysRevD.101.023529}{{\em Phys. Rev. D}
  {\bfseries 101} no.~2, (2020) 023529},
  \href{http://arxiv.org/abs/1912.05927}{{\ttfamily arXiv:1912.05927
  [astro-ph.CO]}}.

\bibitem{Gao:2021vxb}
Q.~Gao, ``{Primordial black holes and secondary gravitational waves from
  chaotic inflation},'' \href{http://dx.doi.org/10.1007/s11433-021-1708-9}{{\em
  Sci. China Phys. Mech. Astron.} {\bfseries 64} no.~8, (2021) 280411},
  \href{http://arxiv.org/abs/2102.07369}{{\ttfamily arXiv:2102.07369 [gr-qc]}}.

\bibitem{Yi:2020cut}
Z.~Yi, Q.~Gao, Y.~Gong, and Z.-h. Zhu, ``{Primordial black holes and
  scalar-induced secondary gravitational waves from inflationary models with a
  noncanonical kinetic term},''
  \href{http://dx.doi.org/10.1103/PhysRevD.103.063534}{{\em Phys. Rev. D}
  {\bfseries 103} no.~6, (2021) 063534},
  \href{http://arxiv.org/abs/2011.10606}{{\ttfamily arXiv:2011.10606
  [astro-ph.CO]}}.

\bibitem{Lin:2021vwc}
J.~Lin, S.~Gao, Y.~Gong, Y.~Lu, Z.~Wang, and F.~Zhang, ``{Primordial black
  holes and scalar induced secondary gravitational waves from Higgs inflation
  with non-canonical kinetic term},''
  \href{http://arxiv.org/abs/2111.01362}{{\ttfamily arXiv:2111.01362 [gr-qc]}}.

\bibitem{Bhaumik:2020dor}
N.~Bhaumik and R.~K. Jain, ``{Small scale induced gravitational waves from
  primordial black holes, a~stringent lower mass bound, and the imprints of an
  early matter to~radiation transition},''
  \href{http://dx.doi.org/10.1103/PhysRevD.104.023531}{{\em Phys. Rev. D}
  {\bfseries 104} no.~2, (2021) 023531},
  \href{http://arxiv.org/abs/2009.10424}{{\ttfamily arXiv:2009.10424
  [astro-ph.CO]}}.

\bibitem{Ragavendra:2020sop}
H.~V. Ragavendra, P.~Saha, L.~Sriramkumar, and J.~Silk, ``{Primordial black
  holes and secondary gravitational waves from ultraslow roll and punctuated
  inflation},'' \href{http://dx.doi.org/10.1103/PhysRevD.103.083510}{{\em Phys.
  Rev. D} {\bfseries 103} no.~8, (2021) 083510},
  \href{http://arxiv.org/abs/2008.12202}{{\ttfamily arXiv:2008.12202
  [astro-ph.CO]}}.

\bibitem{Lin:2020goi}
J.~Lin, Q.~Gao, Y.~Gong, Y.~Lu, C.~Zhang, and F.~Zhang, ``{Primordial black
  holes and secondary gravitational waves from $k$ and $G$ inflation},''
  \href{http://dx.doi.org/10.1103/PhysRevD.101.103515}{{\em Phys. Rev. D}
  {\bfseries 101} no.~10, (2020) 103515},
  \href{http://arxiv.org/abs/2001.05909}{{\ttfamily arXiv:2001.05909 [gr-qc]}}.

\bibitem{2009Natur.460..990A}
B.~P. {Abbott} {\em et~al.}, ``{An upper limit on the stochastic
  gravitational-wave background of cosmological origin},''
  \href{http://dx.doi.org/10.1038/nature08278}{{\em \nat} {\bfseries 460}
  no.~7258, (Aug., 2009) 990--994},
  \href{http://arxiv.org/abs/0910.5772}{{\ttfamily arXiv:0910.5772
  [astro-ph.CO]}}.

\bibitem{Assadullahi:2009jc}
H.~Assadullahi and D.~Wands, ``{Constraints on primordial density perturbations
  from induced gravitational waves},''
  \href{http://dx.doi.org/10.1103/PhysRevD.81.023527}{{\em Phys. Rev. D}
  {\bfseries 81} (2010) 023527},
  \href{http://arxiv.org/abs/0907.4073}{{\ttfamily arXiv:0907.4073
  [astro-ph.CO]}}.

\bibitem{Inomata:2018epa}
K.~Inomata and T.~Nakama, ``{Gravitational waves induced by scalar
  perturbations as probes of the small-scale primordial spectrum},''
  \href{http://dx.doi.org/10.1103/PhysRevD.99.043511}{{\em Phys. Rev. D}
  {\bfseries 99} no.~4, (2019) 043511},
  \href{http://arxiv.org/abs/1812.00674}{{\ttfamily arXiv:1812.00674
  [astro-ph.CO]}}.

\bibitem{Lentati:2015qwp}
L.~Lentati {\em et~al.}, ``{European Pulsar Timing Array Limits On An Isotropic
  Stochastic Gravitational-Wave Background},''
  \href{http://dx.doi.org/10.1093/mnras/stv1538}{{\em Mon. Not. Roy. Astron.
  Soc.} {\bfseries 453} no.~3, (2015) 2576--2598},
  \href{http://arxiv.org/abs/1504.03692}{{\ttfamily arXiv:1504.03692
  [astro-ph.CO]}}.

\bibitem{NANOGrav:2015aud}
{\bfseries NANOGrav} Collaboration, Z.~Arzoumanian {\em et~al.}, ``{The
  NANOGrav Nine-year Data Set: Limits on the Isotropic Stochastic Gravitational
  Wave Background},'' \href{http://dx.doi.org/10.3847/0004-637X/821/1/13}{{\em
  Astrophys. J.} {\bfseries 821} no.~1, (2016) 13},
  \href{http://arxiv.org/abs/1508.03024}{{\ttfamily arXiv:1508.03024
  [astro-ph.GA]}}.

\bibitem{Janssen:2014dka}
G.~Janssen {\em et~al.}, ``{Gravitational wave astronomy with the SKA},''
  \href{http://dx.doi.org/10.22323/1.215.0037}{{\em PoS} {\bfseries AASKA14}
  (2015) 037}, \href{http://arxiv.org/abs/1501.00127}{{\ttfamily
  arXiv:1501.00127 [astro-ph.IM]}}.

\bibitem{KAGRA:2013rdx}
{\bfseries KAGRA, LIGO Scientific, Virgo, VIRGO} Collaboration, B.~P. Abbott
  {\em et~al.}, ``{Prospects for observing and localizing gravitational-wave
  transients with Advanced LIGO, Advanced Virgo and KAGRA},''
  \href{http://dx.doi.org/10.1007/s41114-020-00026-9}{{\em Living Rev. Rel.}
  {\bfseries 21} no.~1, (2018) 3},
  \href{http://arxiv.org/abs/1304.0670}{{\ttfamily arXiv:1304.0670 [gr-qc]}}.

\bibitem{2017arXiv170200786A}
P.~{Amaro-Seoane} {\em et~al.}, ``{Laser Interferometer Space Antenna},'' {\em
  arXiv e-prints} (Feb., 2017) ,
  \href{http://arxiv.org/abs/1702.00786}{{\ttfamily arXiv:1702.00786
  [astro-ph.IM]}}.

\bibitem{Yagi:2011wg}
K.~Yagi and N.~Seto, ``{Detector configuration of DECIGO/BBO and identification
  of cosmological neutron-star binaries},''
  \href{http://dx.doi.org/10.1103/PhysRevD.83.044011}{{\em Phys. Rev. D}
  {\bfseries 83} (2011) 044011},
  \href{http://arxiv.org/abs/1101.3940}{{\ttfamily arXiv:1101.3940
  [astro-ph.CO]}}. [Erratum: Phys.Rev.D 95, 109901 (2017)].

\bibitem{ET}
``{Einstein Telescope webpage}.''. \url{http://www.et-gw.eu/}.

\bibitem{2015PhRvD..91h2001D}
S.~{Dwyer}, D.~{Sigg}, S.~W. {Ballmer}, L.~{Barsotti}, N.~{Mavalvala}, and
  M.~{Evans}, ``{Gravitational wave detector with cosmological reach},''
  \href{http://dx.doi.org/10.1103/PhysRevD.91.082001}{{\em \prd} {\bfseries 91}
  no.~8, (Apr., 2015) 082001}, \href{http://arxiv.org/abs/1410.0612}{{\ttfamily
  arXiv:1410.0612 [astro-ph.IM]}}.

\bibitem{Domenech:2021ztg}
G.~Dom\`enech, ``{Scalar Induced Gravitational Waves Review},''
  \href{http://dx.doi.org/10.3390/universe7110398}{{\em Universe} {\bfseries 7}
  no.~11, (2021) 398}, \href{http://arxiv.org/abs/2109.01398}{{\ttfamily
  arXiv:2109.01398 [gr-qc]}}.

\bibitem{Bian:2021ini}
L.~Bian {\em et~al.}, ``{The Gravitational-wave physics II: Progress},''
  \href{http://dx.doi.org/10.1007/s11433-021-1781-x}{{\em Sci. China Phys.
  Mech. Astron.} {\bfseries 64} no.~12, (2021) 120401},
  \href{http://arxiv.org/abs/2106.10235}{{\ttfamily arXiv:2106.10235 [gr-qc]}}.

\bibitem{Berera:1995wh}
A.~Berera and L.-Z. Fang, ``{Thermally induced density perturbations in the
  inflation era},'' \href{http://dx.doi.org/10.1103/PhysRevLett.74.1912}{{\em
  Phys. Rev. Lett.} {\bfseries 74} (1995) 1912--1915},
\href{http://arxiv.org/abs/astro-ph/9501024}{{\ttfamily arXiv:astro-ph/9501024
  [astro-ph]}}.

\bibitem{Berera:1995ie}
A.~Berera, ``{Warm inflation},''
  \href{http://dx.doi.org/10.1103/PhysRevLett.75.3218}{{\em Phys. Rev. Lett.}
  {\bfseries 75} (1995) 3218--3221},
\href{http://arxiv.org/abs/astro-ph/9509049}{{\ttfamily arXiv:astro-ph/9509049
  [astro-ph]}}.

\bibitem{Berera:1998px}
A.~Berera, M.~Gleiser, and R.~O. Ramos, ``{A First principles warm inflation
  model that solves the cosmological horizon / flatness problems},''
  \href{http://dx.doi.org/10.1103/PhysRevLett.83.264}{{\em Phys. Rev. Lett.}
  {\bfseries 83} (1999) 264--267},
\href{http://arxiv.org/abs/hep-ph/9809583}{{\ttfamily arXiv:hep-ph/9809583
  [hep-ph]}}.

\bibitem{Berera:2006xq}
A.~Berera, ``{The warm inflationary universe},''
  \href{http://dx.doi.org/10.1080/00107510500392030}{{\em Contemp. Phys.}
  {\bfseries 47} (2006) 33--49},
\href{http://arxiv.org/abs/0809.4198}{{\ttfamily arXiv:0809.4198 [hep-ph]}}.

\bibitem{Berera:2008ar}
A.~Berera, I.~G. Moss, and R.~O. Ramos, ``{Warm Inflation and its Microphysical
  Basis},'' \href{http://dx.doi.org/10.1088/0034-4885/72/2/026901}{{\em Rept.
  Prog. Phys.} {\bfseries 72} (2009) 026901},
\href{http://arxiv.org/abs/0808.1855}{{\ttfamily arXiv:0808.1855 [hep-ph]}}.

\bibitem{Bastero-Gil:2015nja}
M.~Bastero-Gil, A.~Berera, and N.~Kronberg, ``{Exploring the Parameter Space of
  Warm-Inflation Models},''
  \href{http://dx.doi.org/10.1088/1475-7516/2015/12/046}{{\em JCAP} {\bfseries
  1512} no.~12, (2015) 046},
\href{http://arxiv.org/abs/1509.07604}{{\ttfamily arXiv:1509.07604 [hep-ph]}}.

\bibitem{Visinelli:2016rhn}
L.~Visinelli, ``{Observational Constraints on Monomial Warm Inflation},''
  \href{http://dx.doi.org/10.1088/1475-7516/2016/07/054}{{\em JCAP} {\bfseries
  1607} no.~07, (2016) 054},
\href{http://arxiv.org/abs/1605.06449}{{\ttfamily arXiv:1605.06449
  [astro-ph.CO]}}.

\bibitem{Benetti:2016jhf}
M.~Benetti and R.~O. Ramos, ``{Warm inflation dissipative effects: predictions
  and constraints from the Planck data},''
  \href{http://dx.doi.org/10.1103/PhysRevD.95.023517}{{\em Phys. Rev.}
  {\bfseries D95} no.~2, (2017) 023517},
\href{http://arxiv.org/abs/1610.08758}{{\ttfamily arXiv:1610.08758
  [astro-ph.CO]}}.

\bibitem{Arya:2017zlb}
R.~Arya, A.~Dasgupta, G.~Goswami, J.~Prasad, and R.~Rangarajan, ``{Revisiting
  CMB constraints on warm inflation},''
  \href{http://dx.doi.org/10.1088/1475-7516/2018/02/043}{{\em JCAP} {\bfseries
  1802} no.~02, (2018) 043},
\href{http://arxiv.org/abs/1710.11109}{{\ttfamily arXiv:1710.11109
  [astro-ph.CO]}}.

\bibitem{Bastero-Gil:2017wwl}
M.~Bastero-Gil, S.~Bhattacharya, K.~Dutta, and M.~R. Gangopadhyay,
  ``{Constraining Warm Inflation with CMB data},''
  \href{http://dx.doi.org/10.1088/1475-7516/2018/02/054}{{\em JCAP} {\bfseries
  1802} no.~02, (2018) 054},
\href{http://arxiv.org/abs/1710.10008}{{\ttfamily arXiv:1710.10008
  [astro-ph.CO]}}.

\bibitem{Arya:2018sgw}
R.~Arya and R.~Rangarajan, ``{Study of warm inflationary models and their
  parameter estimation from CMB},''
  \href{http://arxiv.org/abs/1812.03107}{{\ttfamily arXiv:1812.03107
  [astro-ph.CO]}}.

\bibitem{Bastero-Gil:2021fac}
M.~Bastero-Gil and M.~S. D\'\i{}az-Blanco, ``{Gravity waves and primordial
  black holes in scalar warm little inflation},''
  \href{http://dx.doi.org/10.1088/1475-7516/2021/12/052}{{\em JCAP} {\bfseries
  12} no.~12, (2021) 052}, \href{http://arxiv.org/abs/2105.08045}{{\ttfamily
  arXiv:2105.08045 [hep-ph]}}.

\bibitem{Aggarwal:2020umq}
N.~Aggarwal {\em et~al.}, ``{Searching for New Physics with a
  Levitated-Sensor-Based Gravitational-Wave Detector},''
  \href{http://dx.doi.org/10.1103/PhysRevLett.128.111101}{{\em Phys. Rev.
  Lett.} {\bfseries 128} no.~11, (2022) 111101},
  \href{http://arxiv.org/abs/2010.13157}{{\ttfamily arXiv:2010.13157 [gr-qc]}}.

\bibitem{Bernard:2002ci}
P.~Bernard, A.~Chincarini, G.~Gemme, R.~Parodi, and E.~Picasso, ``{A Detector
  of gravitational waves based on coupled microwave cavities},''
  \href{http://arxiv.org/abs/gr-qc/0203024}{{\ttfamily arXiv:gr-qc/0203024}}.

\bibitem{Holometer:2016qoh}
{\bfseries Holometer} Collaboration, A.~S. Chou {\em et~al.}, ``{MHz
  Gravitational Wave Constraints with Decameter Michelson Interferometers},''
  \href{http://dx.doi.org/10.1103/PhysRevD.95.063002}{{\em Phys. Rev. D}
  {\bfseries 95} no.~6, (2017) 063002},
  \href{http://arxiv.org/abs/1611.05560}{{\ttfamily arXiv:1611.05560
  [astro-ph.IM]}}.

\bibitem{Aguiar:2010kn}
O.~D. Aguiar, ``{The Past, Present and Future of the Resonant-Mass
  Gravitational Wave Detectors},''
  \href{http://dx.doi.org/10.1088/1674-4527/11/1/001}{{\em Res. Astron.
  Astrophys.} {\bfseries 11} (2011) 1--42},
  \href{http://arxiv.org/abs/1009.1138}{{\ttfamily arXiv:1009.1138
  [astro-ph.IM]}}.

\bibitem{Berera:1996fm}
A.~Berera, ``{Interpolating the stage of exponential expansion in the early
  universe: A Possible alternative with no reheating},''
  \href{http://dx.doi.org/10.1103/PhysRevD.55.3346}{{\em Phys. Rev. D}
  {\bfseries 55} (1997) 3346--3357},
  \href{http://arxiv.org/abs/hep-ph/9612239}{{\ttfamily arXiv:hep-ph/9612239}}.

\bibitem{Gupta:2002kn}
S.~Gupta, A.~Berera, A.~F. Heavens, and S.~Matarrese, ``{Non-Gaussian
  signatures in the cosmic background radiation from warm inflation},''
  \href{http://dx.doi.org/10.1103/PhysRevD.66.043510}{{\em Phys. Rev.}
  {\bfseries D66} (2002) 043510},
\href{http://arxiv.org/abs/astro-ph/0205152}{{\ttfamily arXiv:astro-ph/0205152
  [astro-ph]}}.

\bibitem{Moss:2011qc}
I.~G. Moss and T.~Yeomans, ``{Non-gaussianity in the strong regime of warm
  inflation},'' \href{http://dx.doi.org/10.1088/1475-7516/2011/08/009}{{\em
  JCAP} {\bfseries 1108} (2011) 009},
\href{http://arxiv.org/abs/1102.2833}{{\ttfamily arXiv:1102.2833
  [astro-ph.CO]}}.

\bibitem{Berera:2004vm}
A.~Berera, ``{Warm inflation solution to the eta problem},''
  \href{http://arxiv.org/abs/hep-ph/0401139}{{\ttfamily arXiv:hep-ph/0401139
  [hep-ph]}}.
[PoSAHEP2003,069(2003)].

\bibitem{Das:2018rpg}
S.~Das, ``{Warm Inflation in the light of Swampland Criteria},''
  \href{http://dx.doi.org/10.1103/PhysRevD.99.063514}{{\em Phys. Rev.}
  {\bfseries D99} no.~6, (2019) 063514},
\href{http://arxiv.org/abs/1810.05038}{{\ttfamily arXiv:1810.05038 [hep-th]}}.

\bibitem{Motaharfar:2018zyb}
M.~Motaharfar, V.~Kamali, and R.~O. Ramos, ``{Warm inflation as a way out of
  the swampland},'' \href{http://dx.doi.org/10.1103/PhysRevD.99.063513}{{\em
  Phys. Rev.} {\bfseries D99} no.~6, (2019) 063513},
\href{http://arxiv.org/abs/1810.02816}{{\ttfamily arXiv:1810.02816
  [astro-ph.CO]}}.

\bibitem{PhysRevD.37.2878}
E.~Calzetta and B.~L. Hu, ``Nonequilibrium quantum fields: Closed-time-path
  effective action, wigner function, and boltzmann equation,''
  \href{http://dx.doi.org/10.1103/PhysRevD.37.2878}{{\em Phys. Rev. D}
  {\bfseries 37} (May, 1988) 2878--2900}.
  \url{https://link.aps.org/doi/10.1103/PhysRevD.37.2878}.

\bibitem{Gleiser:1993ea}
M.~Gleiser and R.~O. Ramos, ``{Microphysical approach to nonequilibrium
  dynamics of quantum fields},''
  \href{http://dx.doi.org/10.1103/PhysRevD.50.2441}{{\em Phys. Rev.} {\bfseries
  D50} (1994) 2441--2455},
\href{http://arxiv.org/abs/hep-ph/9311278}{{\ttfamily arXiv:hep-ph/9311278
  [hep-ph]}}.

\bibitem{Das:1997gg}
A.~K. Das, {\em {Finite Temperature Field Theory}}.
\newblock World Scientific, New York, 1997.

\bibitem{lebellac:1996}
M.~Le~Bellac, \href{http://dx.doi.org/10.1017/CBO9780511721700}{{\em {Thermal
  Field Theory}}}.
\newblock Cambridge Monographs on Mathematical Physics. Cambridge University
  Press, 1996.

\bibitem{Berera:1998gx}
A.~Berera, M.~Gleiser, and R.~O. Ramos, ``{Strong dissipative behavior in
  quantum field theory},''
  \href{http://dx.doi.org/10.1103/PhysRevD.58.123508}{{\em Phys. Rev. D}
  {\bfseries 58} (1998) 123508},
  \href{http://arxiv.org/abs/hep-ph/9803394}{{\ttfamily arXiv:hep-ph/9803394}}.

\bibitem{Hall:2003zp}
L.~M.~H. Hall, I.~G. Moss, and A.~Berera, ``{Scalar perturbation spectra from
  warm inflation},'' \href{http://dx.doi.org/10.1103/PhysRevD.69.083525}{{\em
  Phys. Rev.} {\bfseries D69} (2004) 083525},
\href{http://arxiv.org/abs/astro-ph/0305015}{{\ttfamily arXiv:astro-ph/0305015
  [astro-ph]}}.

\bibitem{Moss:2008yb}
I.~G. Moss and C.~Xiong, ``{On the consistency of warm inflation},''
  \href{http://dx.doi.org/10.1088/1475-7516/2008/11/023}{{\em JCAP} {\bfseries
  0811} (2008) 023},
\href{http://arxiv.org/abs/0808.0261}{{\ttfamily arXiv:0808.0261 [astro-ph]}}.

\bibitem{Yokoyama:1998ju}
J.~Yokoyama and A.~D. Linde, ``{Is warm inflation possible?},''
  \href{http://dx.doi.org/10.1103/PhysRevD.60.083509}{{\em Phys. Rev. D}
  {\bfseries 60} (1999) 083509},
  \href{http://arxiv.org/abs/hep-ph/9809409}{{\ttfamily arXiv:hep-ph/9809409}}.

\bibitem{Bastero-Gil:2018yen}
M.~Bastero-Gil, A.~Berera, R.~Hernández-Jiménez, and J.~G. Rosa, ``{Warm
  inflation within a supersymmetric distributed mass model},''
  \href{http://dx.doi.org/10.1103/PhysRevD.99.103520}{{\em Phys. Rev.}
  {\bfseries D99} no.~10, (2019) 103520},
\href{http://arxiv.org/abs/1812.07296}{{\ttfamily arXiv:1812.07296 [hep-ph]}}.

\bibitem{Berera:2001gs}
A.~Berera and R.~O. Ramos, ``{The Affinity for scalar fields to dissipate},''
  \href{http://dx.doi.org/10.1103/PhysRevD.63.103509}{{\em Phys. Rev.}
  {\bfseries D63} (2001) 103509},
\href{http://arxiv.org/abs/hep-ph/0101049}{{\ttfamily arXiv:hep-ph/0101049
  [hep-ph]}}.

\bibitem{Berera:2004kc}
A.~Berera and R.~O. Ramos, ``{Dynamics of interacting scalar fields in
  expanding space-time},''
  \href{http://dx.doi.org/10.1103/PhysRevD.71.023513}{{\em Phys. Rev.}
  {\bfseries D71} (2005) 023513},
\href{http://arxiv.org/abs/hep-ph/0406339}{{\ttfamily arXiv:hep-ph/0406339
  [hep-ph]}}.

\bibitem{Bastero-Gil:2016qru}
M.~Bastero-Gil, A.~Berera, R.~O. Ramos, and J.~G. Rosa, ``{Warm Little
  Inflaton},'' \href{http://dx.doi.org/10.1103/PhysRevLett.117.151301}{{\em
  Phys. Rev. Lett.} {\bfseries 117} no.~15, (2016) 151301},
\href{http://arxiv.org/abs/1604.08838}{{\ttfamily arXiv:1604.08838 [hep-ph]}}.

\bibitem{BASTEROGIL2021136055}
M.~Bastero-Gil, A.~Berera, R.~O. Ramos, and J.~G. Rosa, ``Towards a reliable
  effective field theory of inflation,''
  \href{http://dx.doi.org/https://doi.org/10.1016/j.physletb.2020.136055}{{\em
  Physics Letters B} {\bfseries 813} (2021) 136055}.

\bibitem{Moss:2006gt}
I.~G. Moss and C.~Xiong, ``{Dissipation coefficients for supersymmetric
  inflatonary models},''
\href{http://arxiv.org/abs/hep-ph/0603266}{{\ttfamily arXiv:hep-ph/0603266
  [hep-ph]}}.

\bibitem{BasteroGil:2010pb}
M.~Bastero-Gil, A.~Berera, and R.~O. Ramos, ``{Dissipation coefficients from
  scalar and fermion quantum field interactions},''
  \href{http://dx.doi.org/10.1088/1475-7516/2011/09/033}{{\em JCAP} {\bfseries
  1109} (2011) 033},
\href{http://arxiv.org/abs/1008.1929}{{\ttfamily arXiv:1008.1929 [hep-ph]}}.

\bibitem{BasteroGil:2012cm}
M.~Bastero-Gil, A.~Berera, R.~O. Ramos, and J.~G. Rosa, ``{General dissipation
  coefficient in low-temperature warm inflation},''
  \href{http://dx.doi.org/10.1088/1475-7516/2013/01/016}{{\em JCAP} {\bfseries
  1301} (2013) 016},
\href{http://arxiv.org/abs/1207.0445}{{\ttfamily arXiv:1207.0445 [hep-ph]}}.

\bibitem{Graham:2009bf}
C.~Graham and I.~G. Moss, ``{Density fluctuations from warm inflation},''
  \href{http://dx.doi.org/10.1088/1475-7516/2009/07/013}{{\em JCAP} {\bfseries
  0907} (2009) 013},
\href{http://arxiv.org/abs/0905.3500}{{\ttfamily arXiv:0905.3500
  [astro-ph.CO]}}.

\bibitem{BasteroGil:2011xd}
M.~Bastero-Gil, A.~Berera, and R.~O. Ramos, ``{Shear viscous effects on the
  primordial power spectrum from warm inflation},''
  \href{http://dx.doi.org/10.1088/1475-7516/2011/07/030}{{\em JCAP} {\bfseries
  1107} (2011) 030},
\href{http://arxiv.org/abs/1106.0701}{{\ttfamily arXiv:1106.0701
  [astro-ph.CO]}}.

\bibitem{Ramos:2013nsa}
R.~O. Ramos and L.~A. da~Silva, ``{Power spectrum for inflation models with
  quantum and thermal noises},''
  \href{http://dx.doi.org/10.1088/1475-7516/2013/03/032}{{\em JCAP} {\bfseries
  1303} (2013) 032},
\href{http://arxiv.org/abs/1302.3544}{{\ttfamily arXiv:1302.3544
  [astro-ph.CO]}}.

\bibitem{Bartrum:2013fia}
S.~Bartrum, M.~Bastero-Gil, A.~Berera, R.~Cerezo, R.~O. Ramos, and J.~G. Rosa,
  ``{The importance of being warm (during inflation)},''
  \href{http://dx.doi.org/10.1016/j.physletb.2014.03.029}{{\em Phys. Lett.}
  {\bfseries B732} (2014) 116--121},
\href{http://arxiv.org/abs/1307.5868}{{\ttfamily arXiv:1307.5868 [hep-ph]}}.

\bibitem{Kohri:2018awv}
K.~Kohri and T.~Terada, ``{Semianalytic calculation of gravitational wave
  spectrum nonlinearly induced from primordial curvature perturbations},''
  \href{http://dx.doi.org/10.1103/PhysRevD.97.123532}{{\em Phys. Rev. D}
  {\bfseries 97} no.~12, (2018) 123532},
  \href{http://arxiv.org/abs/1804.08577}{{\ttfamily arXiv:1804.08577 [gr-qc]}}.

\bibitem{Maggiore:1999vm}
M.~Maggiore, ``{Gravitational wave experiments and early universe cosmology},''
  \href{http://dx.doi.org/10.1016/S0370-1573(99)00102-7}{{\em Phys. Rept.}
  {\bfseries 331} (2000) 283--367},
  \href{http://arxiv.org/abs/gr-qc/9909001}{{\ttfamily arXiv:gr-qc/9909001}}.

\bibitem{Moore:2014lga}
C.~J. Moore, R.~H. Cole, and C.~P.~L. Berry, ``{Gravitational-wave sensitivity
  curves},'' \href{http://dx.doi.org/10.1088/0264-9381/32/1/015014}{{\em Class.
  Quant. Grav.} {\bfseries 32} no.~1, (2015) 015014},
  \href{http://arxiv.org/abs/1408.0740}{{\ttfamily arXiv:1408.0740 [gr-qc]}}.

\bibitem{Mandic}
V.~Mandic and E.~Floden, ``{GW plotter webpage}.''.
  \url{https://homepages.spa.umn.edu/gwplotter}.

\bibitem{LIGOScientific:2006zmq}
{\bfseries LIGO Scientific} Collaboration, B.~Abbott {\em et~al.}, ``{Searching
  for a Stochastic Background of Gravitational Waves with LIGO},''
  \href{http://dx.doi.org/10.1086/511329}{{\em Astrophys. J.} {\bfseries 659}
  (2007) 918--930}, \href{http://arxiv.org/abs/astro-ph/0608606}{{\ttfamily
  arXiv:astro-ph/0608606}}.

\bibitem{Sendra:2012wh}
I.~Sendra and T.~L. Smith, ``{Improved limits on short-wavelength gravitational
  waves from the cosmic microwave background},''
  \href{http://dx.doi.org/10.1103/PhysRevD.85.123002}{{\em Phys. Rev. D}
  {\bfseries 85} (2012) 123002},
  \href{http://arxiv.org/abs/1203.4232}{{\ttfamily arXiv:1203.4232
  [astro-ph.CO]}}.

\bibitem{Smith:2006nka}
T.~L. Smith, E.~Pierpaoli, and M.~Kamionkowski, ``{A new cosmic microwave
  background constraint to primordial gravitational waves},''
  \href{http://dx.doi.org/10.1103/PhysRevLett.97.021301}{{\em Phys. Rev. Lett.}
  {\bfseries 97} (2006) 021301},
  \href{http://arxiv.org/abs/astro-ph/0603144}{{\ttfamily
  arXiv:astro-ph/0603144}}.

\bibitem{Pagano:2015hma}
L.~Pagano, L.~Salvati, and A.~Melchiorri, ``{New constraints on primordial
  gravitational waves from Planck 2015},''
  \href{http://dx.doi.org/10.1016/j.physletb.2016.07.078}{{\em Phys. Lett. B}
  {\bfseries 760} (2016) 823--825},
  \href{http://arxiv.org/abs/1508.02393}{{\ttfamily arXiv:1508.02393
  [astro-ph.CO]}}.

\bibitem{Parikh:2020fhy}
M.~Parikh, F.~Wilczek, and G.~Zahariade, ``{Signatures of the quantization of
  gravity at gravitational wave detectors},''
  \href{http://dx.doi.org/10.1103/PhysRevD.104.046021}{{\em Phys. Rev. D}
  {\bfseries 104} no.~4, (2021) 046021},
  \href{http://arxiv.org/abs/2010.08208}{{\ttfamily arXiv:2010.08208
  [hep-th]}}.

\bibitem{Ligoandvoyager:2022}
V.~Mandic and E.~Floden, ``{LIGO Document Control Center Portal}.''.
  \url{https://dcc.ligo.org/login/index.shtml?entityID=https%3A%2F%2Fdcc.ligo.org%2Fshibboleth-sp&return=https%3A%2F%2Fdcc.ligo.org%2FShibboleth.sso%2FLogin%3FSAMLDS%3D1%26target%3Dss%253Amem%253Ab56ec8609becaa7774c5954273c049e9eebdb089}.

\bibitem{Hogan:2001jn}
C.~J. Hogan and P.~L. Bender, ``{Estimating stochastic gravitational wave
  backgrounds with Sagnac calibration},''
  \href{http://dx.doi.org/10.1103/PhysRevD.64.062002}{{\em Phys. Rev. D}
  {\bfseries 64} (2001) 062002},
  \href{http://arxiv.org/abs/astro-ph/0104266}{{\ttfamily
  arXiv:astro-ph/0104266}}.

\bibitem{Cornish:2001bb}
N.~J. Cornish, ``{Detecting a stochastic gravitational wave background with the
  Laser Interferometer Space Antenna},''
  \href{http://dx.doi.org/10.1103/PhysRevD.65.022004}{{\em Phys. Rev. D}
  {\bfseries 65} (2002) 022004},
  \href{http://arxiv.org/abs/gr-qc/0106058}{{\ttfamily arXiv:gr-qc/0106058}}.

\bibitem{Corbin:2005ny}
V.~Corbin and N.~J. Cornish, ``{Detecting the cosmic gravitational wave
  background with the big bang observer},''
  \href{http://dx.doi.org/10.1088/0264-9381/23/7/014}{{\em Class. Quant. Grav.}
  {\bfseries 23} (2006) 2435--2446},
  \href{http://arxiv.org/abs/gr-qc/0512039}{{\ttfamily arXiv:gr-qc/0512039}}.

\bibitem{Sato:2017dkf}
S.~Sato {\em et~al.}, ``{The status of DECIGO},''
  \href{http://dx.doi.org/10.1088/1742-6596/840/1/012010}{{\em J. Phys. Conf.
  Ser.} {\bfseries 840} no.~1, (2017) 012010}.

\bibitem{Jenet:2006sv}
F.~A. Jenet {\em et~al.}, ``{Upper bounds on the low-frequency stochastic
  gravitational wave background from pulsar timing observations: Current limits
  and future prospects},'' \href{http://dx.doi.org/10.1086/508702}{{\em
  Astrophys. J.} {\bfseries 653} (2006) 1571--1576},
  \href{http://arxiv.org/abs/astro-ph/0609013}{{\ttfamily
  arXiv:astro-ph/0609013}}.

\bibitem{Aggarwal:2020olq}
N.~Aggarwal {\em et~al.}, ``{Challenges and opportunities of gravitational-wave
  searches at MHz to GHz frequencies},''
  \href{http://dx.doi.org/10.1007/s41114-021-00032-5}{{\em Living Rev. Rel.}
  {\bfseries 24} no.~1, (2021) 4},
  \href{http://arxiv.org/abs/2011.12414}{{\ttfamily arXiv:2011.12414 [gr-qc]}}.

\bibitem{Berlin:2021txa}
A.~Berlin {\em et~al.}, ``{Detecting High-Frequency Gravitational Waves with
  Microwave Cavities},'' \href{http://arxiv.org/abs/2112.11465}{{\ttfamily
  arXiv:2112.11465 [hep-ph]}}.

\bibitem{lec}
L.~notes on Non-equilibrium Quantum Field~Theory and cosmological applications.
\newblock \url{http://gravitation.web.ua.pt/jrosa}.

\bibitem{Das:2020xmh}
S.~Das and R.~O. Ramos, ``{Runaway potentials in warm inflation satisfying the
  swampland conjectures},''
  \href{http://dx.doi.org/10.1103/PhysRevD.102.103522}{{\em Phys. Rev. D}
  {\bfseries 102} no.~10, (2020) 103522},
  \href{http://arxiv.org/abs/2007.15268}{{\ttfamily arXiv:2007.15268
  [hep-th]}}.

\bibitem{supp}
S.~Material.
\newblock \url{http://link.aps.org/
  supplemental/10.1103/PhysRevLett.122.161301}.

\end{thebibliography}\endgroup

\end{document}